\documentclass[12pt]{article}  
  
\usepackage{epsfig,epsf}  
\usepackage{graphics}  
\usepackage{cite}  
\usepackage{geometry}
\usepackage{xcolor} 

\usepackage{amsmath}  
\usepackage{amsthm}  
\usepackage{amsfonts}  
\usepackage{amssymb} 
\usepackage{bm} 

\usepackage{slashed}
\usepackage{braket}

\usepackage{hyperref}
\usepackage{longtable}
\usepackage{caption}

\newcommand{\Lbar}{\overline{\Lambda}}

\begin{document}
\allowdisplaybreaks
\thispagestyle{empty}  
  
\begin{flushright}  
{\small  
IPPP/17/65 \\[0.1cm]
\today
}  
\end{flushright}  
  
\vskip1.5cm  
\begin{center}  
\textbf{\Large\boldmath Dimension-six matrix elements for meson mixing and 
lifetimes from sum rules}  
\\  
\vspace{1.5cm}  
{\sc M.~Kirk}, {\sc A.~Lenz} and {\sc T.~Rauh}\\[0.5cm]  
\vspace*{0.5cm} {\it 
IPPP, Department of Physics,
University of Durham,\\
DH1 3LE, United Kingdom}
  
\def\thefootnote{\arabic{footnote}}  
\setcounter{footnote}{0}  
  
\vskip3cm  
\textbf{Abstract}\\  
\vspace{1\baselineskip}  
\parbox{0.9\textwidth}{
The hadronic matrix elements of dimension-six $\Delta F=0,2$ operators 
are crucial inputs for the theory predictions of mixing observables 
and lifetime ratios in the $B$ and $D$ system. We determine them 
using HQET sum rules for three-point correlators. The results of the 
required three-loop computation of the correlators and the one-loop 
computation of the QCD-HQET matching are given in analytic form.
For mixing matrix elements we find very good agreement with recent 
lattice results and comparable theoretical uncertainties. 
For lifetime matrix elements we present the first ever determination
in the $D$ meson sector and the first determination of 
$\Delta B=0$ matrix elements with uncertainties under control -
superseeding preliminary lattice studies stemming from 2001 and earlier.
With our state-of-the-art determination of the bag parameters we predict:
$\tau(B^+)/\tau(B_d^0) = 1.082_{-0.026}^{+0.022}$, 
$\tau(B_s^0)/\tau(B_d^0) = 1.0007\pm0.0025$,
$\tau(D^+)/\tau(D^0) = 2.7_{-0.8}^{+0.7}$
and the mixing-observables in the $B_s$ and $B_d$ system, 
in good agreement with the most recent experimental averages. 
}  
  
\end{center}  
  
  
\newpage  
\setcounter{page}{1}


\section{Introduction\label{sec:intro}}

The mixing of neutral mesons proceeds through flavour-changing neutral 
currents and is therefore loop suppressed in the Standard Model. 
Thus, mixing observables are very sensitive to new physics effects. 
Our ability to constrain new contributions strongly relies on a high 
degree of precision in both experiment and theory. 
Mixing is most pronounced in the $B_s$ system where the relative decay rate 
difference amounts to about 13\%. Here the experimental precision has 
surpassed the theoretical one by a significant margin~\cite{Artuso:2015swg}. 

The theory expression for mixing observables is a product of perturbative
coefficients and non-perturbative matrix elements. The perturbative part 
is known up to NLO-QCD (see the discussion below) and first steps
in the direction of a NNLO-QCD evaluation have recently been performed 
by \cite{Asatrian:2017qaz}.
However, the dominant theoretical uncertainties still stem from
hadronic matrix elements of local $\Delta B = 2$ four-quark operators. 
They are usually determined by lattice 
simulations and results for the leading dimension-six operators are 
available from several collaborations~\cite{Dalgic:2006gp,Carrasco:2013zta,Bazavov:2016nty}. 
If only the latest lattice results~\cite{Bazavov:2016nty} are used, 
small tensions at the level of two sigma emerge in $B_s$ mixing~\cite{Bazavov:2016nty,Jubb:2016mvq}. 
To either settle or solidify this issue, an independent determination of 
the matrix elements and further scrutinization of the theoretical 
methods are necessary. We address both these points in this paper. 

An alternative way to determine hadronic matrix elements is given 
by QCD sum rules~\cite{Shifman:1978bx,Shifman:1978by}. This approach 
employs quark-hadron duality and the analyticity of Green functions 
instead of the discretization of space-time. Thus, its sources of 
uncertainties are entirely different from lattice simulations and 
sum rule analyses can provide truly independent results. 
We determine the hadronic matrix elements of the dimension-six $\Delta B=2$ 
operators for $B$-mixing from a sum rule for three-point correlators 
first introduced in~\cite{Chetyrkin:1985vj}. The sum rule is valid at 
scales $\mu_\rho \sim 1.5\text{ GeV}$ which are much smaller than the 
bottom-quark mass. Therefore the sum rule is formulated in HQET, where 
quantum fluctuations with a characteristic scale of the order of the 
bottom-quark mass have been integrated out. We then run the HQET matrix 
elements up to a scale $\mu_m$ of the order of the bottom-quark mass 
where the matching to QCD can be performed without introducing large 
logarithms. Earlier sum rule results are available for the SM operator 
$Q_1$~\cite{Korner:2003zk,Grozin:2016uqy} and condensate corrections 
have been computed for dimension-six~\cite{Grozin:2016uqy,Mannel:2007am,Mannel:2011zza,Baek:1998vk,Cheng:1998ia} 
and seven~\cite{Mannel:2007am,Mannel:2011zza} operators. 
The same strategy is then applied to determine the matrix elements of 
dimension-six $\Delta B=0$ operators, which are the non-perturbative input 
for calculating ratios of lifetimes of different mesons, like 
$\tau (B^+) / \tau(B_d)$ and $\tau (B_s) / \tau(B_d)$, see
e.g. \cite{Lenz:2015dra} for a review.
Here the perturbative part of the prediction is also known to NLO-QCD.

The theory prediction for the $B_s$ decay rate difference $\Delta \Gamma_s$ 
and for ratios of lifetimes of different $B$ mesons
is based on the Heavy Quark Expansion 
(HQE)~\cite{Khoze:1983yp,Shifman:1984wx,Bigi:1991ir,Bigi:1992su}. 
The HQE is an OPE in the Minkowski domain which has fuelled 
speculations about large violations of duality, in particular for 
$\Delta\Gamma_s$ which is dominated by the $b\to c\bar{c}s$ transition.\footnote{
Interestingly we find that the HQE prediction for the $b\to c\bar{c}s$ 
branching ratio \cite{Krinner:2013cja} is in excellent agreement with 
experiment.} 
A recent confrontation of HQE predictions with experiment has ruled out 
duality violations larger than about~20\%~\cite{Jubb:2016mvq}.
Ratios of meson lifetimes are a good testing ground for the validity of 
the HQE, but have suffered from large hadronic uncertainties~\cite{Lenz:2015dra}
in the past because only outdated lattice results~\cite{DiPierro:1998ty,Becirevic:2001fy} 
for the required $\Delta B = 0$ matrix elements of four quark operators 
were available. We present the first state-of-the-art calculation 
of the  $\Delta B=0$ matrix elements and determine the lifetimes with significantly 
reduced uncertainties. 

In the charm sector the validity of the HQE is rather uncertain due to 
its smaller mass $m_c\sim m_b/3$. The direct translation of the predictions 
for $B$ mixing fails by several orders of magnitude~\cite{Bobrowski:2010xg}. 
However it has been argued that higher-dimensional contributions can lift 
the severe GIM suppression in the charm sector and potentially explain the 
size of mixing observables~\cite{Bigi:2000wn,Georgi:1992as,Ohl:1992sr,Bobrowski:2010xg,Bobrowski:2012jf}. 
$D$-meson lifetimes have been studied recently~\cite{Lenz:2013aua} and have 
shown no indications for a breakdown of the HQE, albeit with large hadronic 
uncertainties. We translate our sum rule results to the charm sector 
as well. The $\Delta C=2$ matrix elements show good agreement with 
lattice results~\cite{Carrasco:2014uya,Carrasco:2015pra,Bazavov:2017weg} 
and the $\Delta C=0$ results are used to update the $D^+-D^0$ lifetime ratio. 

The outline of this work is as follows: In Section~\ref{sec:matching} 
we describe the details of the QCD-HQET matching computation focussing 
on $\Delta B=2$ operators. The sum rule and the calculation of the 
three-point correlators are discussed in Section~\ref{sec:sum_rule}. 
Our results for the matrix elements are presented in 
Section~\ref{sec:results} and compared to other recent works. 
In Section~\ref{sec:lifetimes} we study $\Delta B=0$ operators and 
ratios of $B$-meson lifetimes. We determine the matrix elements of 
$\Delta C=0,2$ operators in Section~\ref{sec:charm} and update the 
HQE result for the $D^+-D^0$ lifetime ratio using these results. 
Finally, we conclude in Section~\ref{sec:conclusion}.


\section{\boldmath QCD-HQET matching for $\Delta B=2$ operators\label{sec:matching}}

We perform the matching computation between QCD and HQET operators 
at the one-loop level. The details of the computation are described 
in Section~\ref{sec:matching_setup} for the $\Delta B=2$ operators. 
Our results for the matching of the operators and Bag parameters 
are given in Section~\ref{matching_results} and Section~\ref{matching_bags}, 
respectively. 

\subsection{Setup\label{sec:matching_setup}}

The matching calculation for the SM operator $Q_1$ appearing in $\Delta M_s$ 
has been performed in~\cite{Flynn:1990qz,Buchalla:1996ys,Ciuchini:1996sr}. 
We compute the matching coefficients of the full dimension-six $\Delta B=2$ 
operator basis needed for $\Delta M_s$ in BSM theories and for $\Delta\Gamma_s$ 
in the SM. We work in dimensional regularization with $d=4-2\epsilon$ and an 
anticommuting $\gamma^5$ (NDR scheme). We consider the following operators in QCD 
\begin{eqnarray}
 Q_1 & = & \bar{b}_i\gamma_\mu(1-\gamma^5)q_i\,\,\bar{b}_j\gamma^\mu(1-\gamma^5)q_j, \nonumber\\
 Q_2 & = & \bar{b}_i(1-\gamma^5)q_i\,\,\bar{b}_j(1-\gamma^5)q_j, \hspace{1cm} Q_3 = \bar{b}_i(1-\gamma^5)q_j\,\,\bar{b}_j(1-\gamma^5)q_i,\nonumber\\
 Q_4 & = & \bar{b}_i(1-\gamma^5)q_i\,\,\bar{b}_j(1+\gamma^5)q_j, \hspace{1cm} Q_5 = \bar{b}_i(1-\gamma^5)q_j\,\,\bar{b}_j(1+\gamma^5)q_i. 
 \label{eq:QCD_operators}
\end{eqnarray}
To fix the renormalization scheme we also have to specify a basis of 
evanescent operators~\cite{Collins:1984xc,Buras:1989xd,Herrlich:1994kh}. 
We do this following~\cite{Beneke:1998sy}. 
The explicit form of the evanescent operators can be found in 
Appendix~\ref{sec:EvOps_and_ADMs}.
On the HQET side, we have the operators 
\begin{eqnarray}
 \tilde{Q}_1 & = & \bar{h}_i^{\{(+)}\gamma_\mu(1-\gamma^5)q_i\,\,\bar{h}_j^{(-)\}}\gamma^\mu(1-\gamma^5)q_j, \hspace{0.5cm}
 \tilde{Q}_2 = \bar{h}_i^{\{(+)}(1-\gamma^5)q_i\,\,\bar{h}_j^{(-)\}}(1-\gamma^5)q_j,\nonumber\\
 \tilde{Q}_4 & = & \bar{h}_i^{\{(+)}(1-\gamma^5)q_i\,\,\bar{h}_j^{(-)\}}(1+\gamma^5)q_j, \hspace{1.35cm} \tilde{Q}_5 = \bar{h}_i^{\{(+)}(1-\gamma^5)q_j\,\,\bar{h}_j^{(-)\}}(1+\gamma^5)q_i,
 \nonumber
 \\
 \label{eq:HQET_operators}
\end{eqnarray}
where the HQET field $h^{(+)}(x)$ annihilates a bottom quark, $h^{(-)}(x)$ 
creates an anti-bottom and we have introduced the notation 
\begin{equation}
 \bar{h}^{\{(+)}\Gamma_A q\,\,\bar{h}^{(-)\}}\Gamma_B q = \bar{h}^{(+)}\Gamma_A q\,\,\bar{h}^{(-)}\Gamma_B q + \bar{h}^{(-)}\Gamma_A q\,\,\bar{h}^{(+)}\Gamma_B q.
\end{equation}
Note that no operator $\tilde{Q}_3$ appears on the HQET side because 
it is not linearly independent, just like its QCD equivalent at leading 
order in $1/m_b$~\cite{Beneke:1996gn}. We define the evanescent HQET 
operators up to three constants $a_i$ with $i=1,2,3$ which allow us to 
keep track of the scheme dependence. Again the explicit basis of the 
evanescent operators can be found in Appendix~\ref{sec:EvOps_and_ADMs}. 
The matching condition for the $\Delta B=2$ operators is given by 
\begin{equation}
 \braket{Q_i}(\mu) = \sum C_{Q_i\tilde{Q}_j}(\mu)\braket{\tilde{Q}_j}(\mu) + \mathcal{O}\left(\frac{1}{m_b}\right),
 \label{eq:matching_condition}
\end{equation}
where $\braket{A}=\braket{\bar{B}|A|B}$. The matching coefficients can 
be expanded in perturbation theory and take the form 
\begin{equation}
 C_{Q_i\tilde{Q}_j}(\mu) = C_{Q_i\tilde{Q}_j}^{(0)} + \frac{\alpha_s(\mu)}{4\pi}C_{Q_i\tilde{Q}_j}^{(1)}(\mu) + \dots \, .
\end{equation}
Thus the matching calculation can be performed with external quark 
states. The partonic QCD matrix elements are 
\begin{equation}
 \braket{Q} = \frac{\delta_{\alpha\beta}\delta_{\gamma\delta}}{N_c} \left[Z_b^\text{OS}Z_q^\text{OS}Z_{Q\mathcal{O}}\left(
 \hspace{-0.7cm}\vcenter{\mbox{\includegraphics[height=2.5cm]{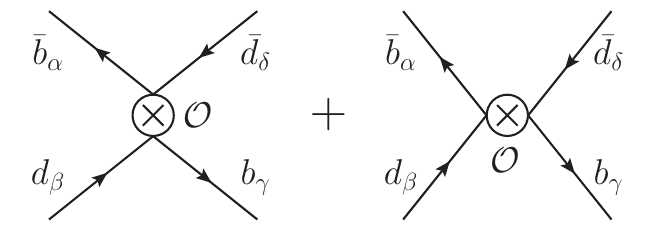}}}\hspace{-7.5cm}\right)+\mathcal{O}(\alpha_s)\right],
\end{equation}
where we sum over $\mathcal{O}$, including all physical and evanescent 
operators, and the color singlet initial and final state have been 
projected out. 
\begin{figure}
 \begin{center}
    \includegraphics[width=\textwidth]{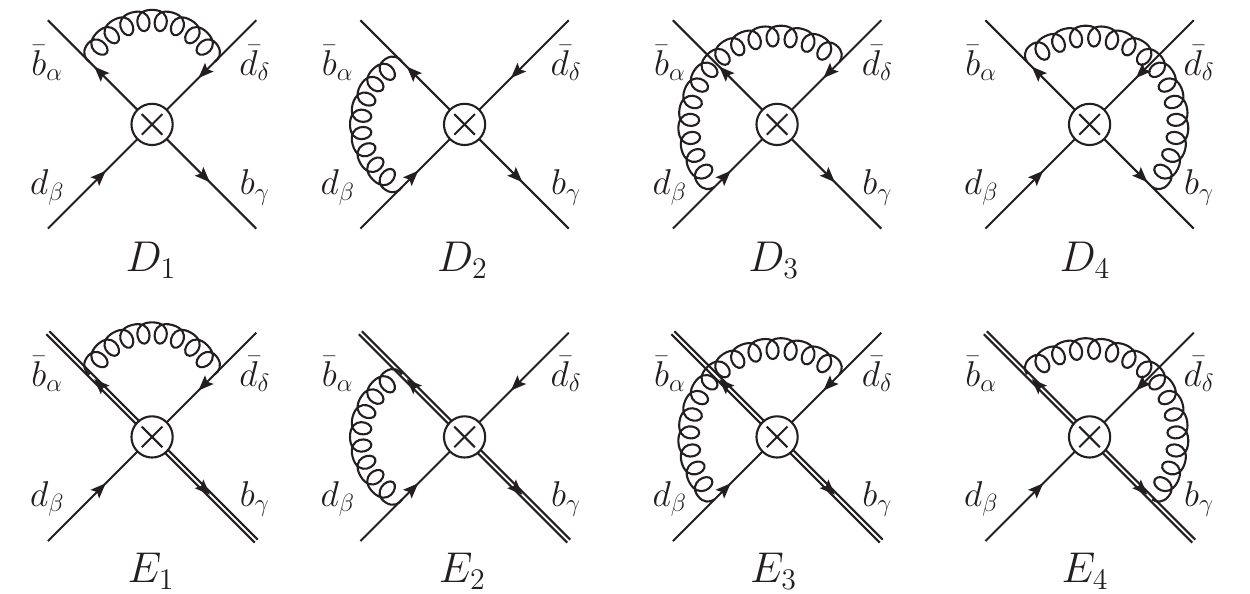}
  \caption{\label{fig:QCD_HQET_matching_diagrams}
  QCD ($D_i$) and HQET ($E_i$) diagrams that enter the matching. Symmetric 
  diagrams are not shown.}
 \end{center}
\end{figure}
The two tree-level contractions appear with a relative minus sign. 
The gluon corrections are shown in Figure~\ref{fig:QCD_HQET_matching_diagrams} 
and do not contain self-energy insertions on the external legs, since 
the quark fields are renormalized in the on-shell scheme. The HQET matrix 
elements follow from the replacements $Q\to\tilde{Q}$, 
$\mathcal{O}\to\tilde{\mathcal{O}}$, $Z_b^\text{OS}\to Z_h^\text{OS}$ and using 
HQET propagators instead of the full QCD ones for the bottom quark. 
The heavy quark on-shell renormalization constants are 
\begin{equation}
 Z_b^\text{OS} = 1 - \frac{\alpha_sC_F}{4\pi}\left(\frac{3}{\epsilon}+4+3\ln\frac{\mu^2}{m_b^2}\right)+\mathcal{O}(\alpha_s^2),\hspace{1cm} Z_h^\text{OS}=1.
\end{equation}
The light-quark renormalization is trivial in the massless case $Z_q^\text{OS}=1$. 
For the renormalization of the physical operators the $\overline{\text{MS}}$ scheme 
is used. In accordance with~\cite{Collins:1984xc,Buras:1989xd,Herrlich:1994kh} the 
evanescent operators are renormalized by a finite amount such that their physical matrix 
elements vanish. Consequently the Wilson coefficients $C_{Q_i\tilde{E}_j}$ 
are not required for the determination of the hadronic matrix elements and are omitted 
in the results shown below. However, in the matching computation itself the matrix elements 
are taken between external on-shell quark states and are therefore not IR finite. 
While the IR divergences cancel in the matching of the QCD to the HQET operators 
there are non-vanishing contributions to the physical matching coefficients 
$C_{Q\tilde{Q}}$ from matrix elements of the evanescent 
operators that are multiplied by IR poles since the evanescent operators are defined 
differently in QCD and HQET, cf. Appendix~\ref{sec:EvOps_and_ADMs}. 

We also find that the NLO matching coefficients $C_{Q_3\tilde{Q}_j}^{(1)}$ of 
the operator $Q_3$ are affected by the finite renormalization of the evanescent 
operator $\tilde{E}_2$ which contains contributions proportional to the physical operators. 
This usually only happens at NNLO (as is the case for the other operators) but is 
already present here at NLO because the tree-level matching coefficient 
$C_{Q_3\tilde{E}_2}^{(0)}$ of this operator is non-vanishing and, therefore, 
the NLO matrix element of the evanescent HQET operator $\tilde{E}_2$ already appears 
at NLO in the matching calculation. 

In the computation we have used both a manual approach and an automated setup 
utilizing \texttt{QGRAF}~\cite{Nogueira:1991ex} and \texttt{Mathematica} to generate 
the amplitudes. The Dirac algebra has been performed with a customized version 
of \texttt{TRACER}~\cite{Jamin:1991dp} as well as with 
\texttt{Package-X}~\cite{Patel:2015tea,Patel:2016fam} and the QCD loop integrals 
have been evaluated using \texttt{Package-X}~\cite{Patel:2015tea,Patel:2016fam}. 
We have also checked our results by performing the calculation with a gluon mass 
as an IR regulator and found full agreement.

\subsection{Results\label{matching_results}}

We write the LO QCD anomalous dimension matrix (ADM) as 
\begin{equation}
 \gamma^{(0)} = \left(\begin{array}{cc}
 \gamma_{QQ}^{(0)} & \gamma_{QE}^{(0)}\\
 \gamma_{EQ}^{(0)} & \gamma_{EE}^{(0)}
\end{array}\right),
\label{eq:ADM_QCD}
\end{equation}
where $\gamma_{QQ}^{(0)}$ is the ADM for the physical set of 
operators~\eqref{eq:QCD_operators}, $\gamma_{QE}^{(0)}$ describes the mixing 
of the physical operators into the evanescent ones~\eqref{eq:QCD_ev_operators}, 
$\gamma_{EQ}^{(0)}$ vanishes (see~\cite{Herrlich:1994kh}) and $\gamma_{EE}^{(0)}$ 
is not required. We decompose the LO HQET ADM $\tilde{\gamma}^{(0)}$ analogously. 
Our results for the non-vanishing entries are given in 
Appendix~\ref{sec:EvOps_and_ADMs}. 

The non-vanishing Wilson coefficients 
at LO are 
\begin{equation}
 C_{Q_1\tilde{Q}_1}^{(0)} = 1, \hspace{0.3cm}
 C_{Q_2\tilde{Q}_2}^{(0)} = 1, \hspace{0.3cm}
 C_{Q_3\tilde{Q}_1}^{(0)} = -\frac12, \hspace{0.3cm}
 C_{Q_3\tilde{Q}_2}^{(0)} = -1, \hspace{0.3cm}
 C_{Q_4\tilde{Q}_4}^{(0)} = 1, \hspace{0.3cm}
 C_{Q_5\tilde{Q}_5}^{(0)} = 1.
\end{equation}
The NLO corrections to the matching coefficients read 
\begin{equation}
 C_{Q\tilde{Q}}^{(1)} = \left(
\begin{array}{cccc}
 -\frac{41}{3}+\frac{a_2}{12}-6 L_\mu \hspace{0.2cm}& -8 \hspace{0.2cm}& 0 \hspace{0.2cm}& 0 \\ \vspace*{0.1cm}
 \frac{3}{2}-\frac{a_1}{12}+L_\mu \hspace{0.2cm}& 8+4 L_\mu \hspace{0.2cm}& 0 \hspace{0.2cm}& 0 \\ \vspace*{0.1cm}
 5+\frac{2a_1-a_2}{24}+4 L_\mu \hspace{0.2cm}& 4+4 L_\mu \hspace{0.2cm}& 0 \hspace{0.2cm}& 0 \\ \vspace*{0.1cm}
 0 \hspace{0.2cm}& 0 \hspace{0.2cm}& 8-\frac{a_3}{24}+\frac{9 L_\mu}{2} \hspace{0.2cm}& -4+\frac{a_3}{8}-\frac{3 L_\mu}{2} \\ \vspace*{0.1cm}
 0 \hspace{0.2cm}& 0 \hspace{0.2cm}& 4+\frac{a_3}{8}+\frac{3 L_\mu}{2} \hspace{0.2cm}& -8-\frac{a_3}{24}-\frac{9 L_\mu}{2} \\
\end{array}
\right),
\label{eq:CQQ}
\end{equation}
where $L_\mu=\ln(\mu^2/m_b^2)$ and we have set $N_c=3$ to keep the 
results compact. 

\subsection{Matching of QCD and HQET Bag parameters\label{matching_bags}}

We define the QCD bag parameters $B_Q$ following~\cite{Gabbiani:1996hi}
\begin{equation}
  \braket{Q(\mu)} = A_Q\, f_B^2M_B^2\, B_Q(\mu),
  \label{eq:Bags_QCD}
\end{equation}
where the coefficients read 
\begin{equation}
 \begin{array}{ll}
  A_{Q_1} = 2+\frac{2}{N_c}, & \\
  A_{Q_2} = \frac{M_B^2}{(m_b+m_q)^2}\left(-2+\frac{1}{N_c}\right), \hspace{1cm}& A_{Q_3} = \frac{M_B^2}{(m_b+m_q)^2}\left(1-\frac{2}{N_c}\right),\\
  A_{Q_4} = \frac{2M_B^2}{(m_b+m_q)^2}+\frac{1}{N_c}, & A_{Q_5} = 1+\frac{2M_B^2}{N_c(m_b+m_q)^2}, 
 \end{array}
 \label{eq:AQi_QCD}
\end{equation}
the $B$ meson decay constant $f_B$ is defined as 
\begin{equation}
 \braket{0|\bar{b}\gamma^\mu\gamma^5 q|B(p)} = -if_Bp^\mu, 
\end{equation}
$M_B$ is the mass of the $B$ meson and $B_{Q_i}=1$ corresponds to the 
VSA approximation. We note that the quark masses appearing in~\eqref{eq:AQi_QCD} 
are \emph{not} $\overline{\text{MS}}$ masses which is the usual convention 
today~\cite{Lenz:2006hd,Bazavov:2016nty}, but pole masses. 
We prefer the definition~\eqref{eq:Bags_QCD} for the analysis because 
the use of $\overline{\text{MS}}$ masses makes the LO ADM of the Bag 
parameters explicitly $\mu$-dependent and prohibits an analytic solution 
of the RGE. At the end we convert our results to the convention 
of~\cite{Lenz:2006hd,Bazavov:2016nty} which we denote as 
\begin{equation}
  \braket{Q(\mu)} = \overline{A}_Q(\mu)\, f_B^2M_B^2\, \overline{B}_Q(\mu),
  \label{eq:Bags_QCD_MSmasses}
\end{equation}
where the $\overline{A}_Q(\mu)$ follow from $A_Q$ with the replacements 
$m_b\to\overline{m}_b(\mu)$ and $m_q\to\overline{m}_q(\mu)$. 
Similar to~\eqref{eq:Bags_QCD}, we use for the HQET operators 
\begin{equation}
 \hm{\langle}\hspace{-0.2cm}\hm{\langle}\tilde{Q}(\mu)\hm{\rangle}\hspace{-0.2cm}\hm{\rangle} = 
 A_{\tilde{Q}}\, F^2(\mu)\, B_{\tilde{Q}}(\mu), 
 \label{eq:Bags_HQET}
\end{equation}
where 
\begin{equation}
  A_{\tilde{Q}_1} = 2+\frac{2}{N_c},\hspace{0.4cm} A_{\tilde{Q}_2} = -2+\frac{1}{N_c},\hspace{0.4cm}
  A_{\tilde{Q}_4} = 2+\frac{1}{N_c},\hspace{0.4cm} A_{\tilde{Q}_5} = 1+\frac{2}{N_c},
 \label{eq:AQi_HQET}
\end{equation}
and the matrix elements have been taken between non-relativistically 
normalized states 
$\hm{\langle}\hspace{-0.2cm}\hm{\langle}\tilde{Q}_i(\mu)\hm{\rangle}\hspace{-0.2cm}\hm{\rangle} 
\equiv \braket{\overline{\mathbf{B}}|\tilde{Q}_i(\mu)|\mathbf{B}}$ 
with 
\begin{equation}
 \Ket{B(p)} = \sqrt{2M_B}\Ket{\mathbf{B}(v)}+\mathcal{O}\left(1/m_b\right),
 \label{eq:Nonrel_states}
\end{equation}
such that 
\begin{equation}
 \braket{\mathbf{B}(v')|\mathbf{B}(v)} = \frac{v^0}{M_B^3}(2\pi)^3\delta^{(3)}(\mathbf{v}'-\mathbf{v}). 
\end{equation}
The parameter $F(\mu)$ is defined as
\begin{equation}
 \braket{0|\bar{h}^{(-)}\gamma^\mu\gamma^5 q|\mathbf{B}(v)} = -iF(\mu)v^\mu, 
\end{equation}
and related to the decay constant by 
\begin{equation}
 f_B = \sqrt{\frac{2}{M_B}}C(\mu)F(\mu)+\mathcal{O}\left(1/m_b\right),
 \label{eq:fB_F}
\end{equation}
with~\cite{Eichten:1989zv} 
\begin{equation}
 C(\mu) = 1-2C_F\frac{\alpha_s(\mu)}{4\pi}+\mathcal{O}(\alpha_s^2).
\end{equation}
From~\eqref{eq:Bags_QCD} and~\eqref{eq:Bags_HQET}, we obtain, using \eqref{eq:matching_condition}, 
\eqref{eq:Nonrel_states} and \eqref{eq:fB_F},
\begin{equation}
 B_{Q_i}(\mu) = \sum\limits_j \frac{A_{\tilde{Q}_j}}{A_{Q_i}}\,\frac{C_{Q_i\tilde{Q}_j}(\mu)}{C^2(\mu)}\,B_{\tilde{Q}_j}(\mu) + \mathcal{O}(1/m_b).
 \label{eq:B_QCD_HQET_matching}
\end{equation}
The HQET bag parameters $B_{\tilde{Q}}$ are determined 
from a sum rule analysis.


\section{HQET sum rule\label{sec:sum_rule}}

The HQET sum rule is introduced in Section~\ref{sec:sum_rule_def}. 
We give results for the double-discontinuity of the three-point 
correlators in Section~\ref{sec:sum_rule_dd} and describe the 
determination of HQET and QCD Bag parameters in Section~\ref{sec:sum_rule_bags}. 

\subsection{The sum rule\label{sec:sum_rule_def}}

We define the three-point correlator 
\begin{equation}
 K_{\tilde{Q}}(\omega_1,\omega_2) = \int d^dx_1 d^dx_2 e^{ip_1\cdot x_1-ip_2\cdot x_2}\braket{0|\text{T}\left[\tilde{j}_{+}(x_2)\tilde{Q}(0)\tilde{j}_{-}(x_1)\right]|0},
 \label{eq:DefK}
\end{equation}
where $\omega_{1,2} = p_{1,2}\cdot v$ and 
\begin{equation}
 \tilde{j}_{+} = \bar{q}\gamma^5 h^{(+)}, \hspace{1cm} \tilde{j}_{-} = \bar{q}\gamma^5 h^{(-)},
\end{equation}
are interpolating currents for the pseudoscalar $\overline{B}$ and $B$ mesons. 
The correlator~\eqref{eq:DefK} is analytic in $\omega_{1,2}$ apart from 
discontinuities for positive real $\omega$. This allows us to construct a 
dispersion relation 
\begin{equation}
 K_{\tilde{Q}}(\omega_1,\omega_2) = \int\limits_{0}^\infty d\eta_1 d\eta_2\,\frac{\rho_{\tilde{Q}}(\eta_1,\eta_2)}{(\eta_1-\omega_1)(\eta_2-\omega_2)}+\left[\text{subtraction terms}\right],
 \label{eq:disp_rel}
\end{equation}
where $\rho_{\tilde{Q}}$ is the \emph{double} discontinuity of $K_{\tilde{Q}}$ 
in $\omega_1$ and $\omega_2$. The second term on the right originates 
from the integration of $K_{\tilde{Q}}$ along the circle at infinity in the 
complex $\eta_1$ or (and) $\eta_2$ planes and is therefore polynomial 
in $\omega_1$ or (and) $\omega_2$. The correlator $K_{\tilde{Q}}$ 
can be computed by means of an OPE 
\begin{equation}
 K_{\tilde{Q}}^{\text{OPE}}(\omega_1,\omega_2) = K_{\tilde{Q}}^{\text{pert}}(\omega_1,\omega_2)
 +K_{\tilde{Q}}^{\braket{\bar{q}q}}(\omega_1,\omega_2)\braket{\bar{q}q}
 +K_{\tilde{Q}}^{\braket{\alpha_sG^2}}(\omega_1,\omega_2)\braket{\alpha_sG^2}+\dots
\end{equation}
for values of $\omega_{1,2}$ that lie far away from the physical cut. 
Assuming quark-hadron duality, we can equate the correlator 
$K_{\tilde{Q}}^{\text{OPE}}$ with its hadronic counterpart 
\begin{equation}
 K_{\tilde{Q}}^{\text{had}}(\omega_1,\omega_2) = \int\limits_{0}^\infty d\eta_1 d\eta_2\,\frac{\rho_{\tilde{Q}}^\text{had}(\eta_1,\eta_2)}{(\eta_1-\omega_1)(\eta_2-\omega_2)}+\left[\text{subtraction terms}\right],
\end{equation}
which is obtained from integration over the hadronic spectral function 
\begin{equation}
 \rho_{\tilde{Q}}^\text{had}(\omega_1,\omega_2) = F^2(\mu)\hm{\langle}\hspace{-0.2cm}\hm{\langle}\tilde{Q}(\mu)\hm{\rangle}\hspace{-0.2cm}\hm{\rangle}
 \delta(\omega_1-\Lbar)\delta(\omega_2-\Lbar) + \rho_{\tilde{Q}}^\text{cont}(\omega_1,\omega_2).
\end{equation}
We use a double Borel transformation with respect to $\omega_{1,2}$ 
to remove the contribution from the integration over the circle at 
infinity and to suppress the sensitivity to the continuum part 
$\rho_{\tilde{Q}}^\text{cont}$ of the spectral function, 
which yields the sum rule 
\begin{equation}
 \int\limits_0^\infty d\omega_1d\omega_2e^{-\frac{\omega_1}{t_1}-\frac{\omega_2}{t_2}}\rho_{\tilde{Q}}^\text{OPE}(\omega_1,\omega_2) = 
 \int\limits_0^\infty d\omega_1d\omega_2e^{-\frac{\omega_1}{t_1}-\frac{\omega_2}{t_2}}\rho_{\tilde{Q}}^\text{had}(\omega_1,\omega_2).
 \label{eq:sum_rule_uncut}
\end{equation}
In principle one can proceed by modelling the continuum $\rho_{\tilde{Q}}^\text{cont}$. 
The desired matrix element of the operator $\tilde{Q}$ between 
the mesonic ground state can then be disentangled by varying the Borel 
parameters. However, the continuum contribution is exponentially suppressed 
in the Borel sum rule and it is safe to simply ``cut off'' the sum rule 
by assuming that 
\begin{equation}
 \rho_{\tilde{Q}}^\text{cont}(\omega_1,\omega_2) = \rho_{\tilde{Q}}^\text{OPE}(\omega_1,\omega_2)\left[1-\theta(\omega_c-\omega_1)\theta(\omega_c-\omega_2)\right],
\end{equation}
which directly yields a finite-energy sum rule for the matrix elements 
\begin{equation}
  F^2(\mu)\hm{\langle}\hspace{-0.2cm}\hm{\langle}\tilde{Q}(\mu)\hm{\rangle}\hspace{-0.2cm}\hm{\rangle} e^{-\frac{\Lbar}{t_1}-\frac{\Lbar}{t_2}} = 
  \int\limits_0^{\omega_c} d\omega_1d\omega_2e^{-\frac{\omega_1}{t_1}-\frac{\omega_2}{t_2}}\rho_{\tilde{Q}}^\text{OPE}(\omega_1,\omega_2).
  \label{eq:SR_matrix_elements}
\end{equation}
Thus, the determination of the HQET Bag parameters requires the computation 
of the spectral functions $\rho_{\tilde{Q}}^\text{OPE}$. The leading 
condensate corrections have been determined in~\cite{Mannel:2007am,Mannel:2011zza}. 
We compute the $\mathcal{O}(\alpha_s)$ corrections to the perturbative 
contribution below.

\subsection{Spectral functions at NLO\label{sec:sum_rule_dd}}

\begin{figure}
 \begin{center}
    \includegraphics[width=0.45\textwidth]{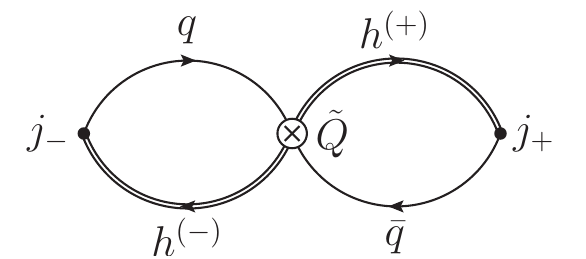}
  \caption{\label{fig:ThreePointCorrelatorLO}
  Leading order diagram for the three-point HQET correlator~\eqref{eq:DefK}. 
  The sum over the two possible contractions of the operator $\tilde{Q}$ 
  is implied.}
 \end{center}
\end{figure}

We determine the spectral functions by first computing the correlator 
\begin{equation}
K_{\tilde{Q}}^\text{pert}(\omega_1,\omega_2) = K_{\tilde{Q}}^{(0)}(\omega_1,\omega_2)+\frac{\alpha_s}{4\pi}K_{\tilde{Q}}^{(1)}(\omega_1,\omega_2)+\dots
\end{equation}
and then taking its double discontinuity. At LO we have to evaluate the diagram 
in Figure~\ref{fig:ThreePointCorrelatorLO} which factorizes into two two-point 
functions. 
We obtain\footnote{As discussed below the sum rule reproduces the 
VSA at LO. Therefore the factors $A_{\tilde{Q}_i}$ appear at leading order 
in the expansion of the results in $\epsilon$. However, the correlator is 
computed in $d$ dimensions and corrections can appear. We find that this happens 
only for $\tilde{Q}_1$ where the contraction of the two $\gamma$ matrices inside 
the trace yields a $d$-dimensional factor.} 
\begin{equation}
 K_{\tilde{Q}_i}^{(0)}(\omega_1,\omega_2) = \left(A_{\tilde{Q}_i} - \delta_{i1}\,\frac{2\epsilon}{N_c}\right) \Pi^{(0)}(\omega_1)\Pi^{(0)}(\omega_2),
 \label{eq:KLO}
\end{equation}
where 
\begin{equation}
 \Pi^{(0)}(\omega) = -\frac{4N_c}{(4\pi)^{2-\epsilon}}\,\tilde{\mu}^{2\epsilon}\,(-2\omega)^{2-2\epsilon}\,\Gamma(2-\epsilon)\Gamma(-2+2\epsilon)
\end{equation}
is the LO result for the two-point correlator 
\begin{equation}
 \Pi(\omega) = i\,\int d^dx e^{ipx}\braket{0|\text{T}\left[\tilde{j}_+^\dagger(0)\tilde{j}_+(x)\right]|0},
 \label{eq:Pi}
\end{equation}
where $\omega=p\cdot v$ and the use of $\tilde{\mu}^2=\mu^2\exp(\gamma_E)/(4\pi)$ 
corresponds to the $\overline{\text{MS}}$ scheme. 

\begin{figure}
 \begin{center}
    \includegraphics[width=\textwidth]{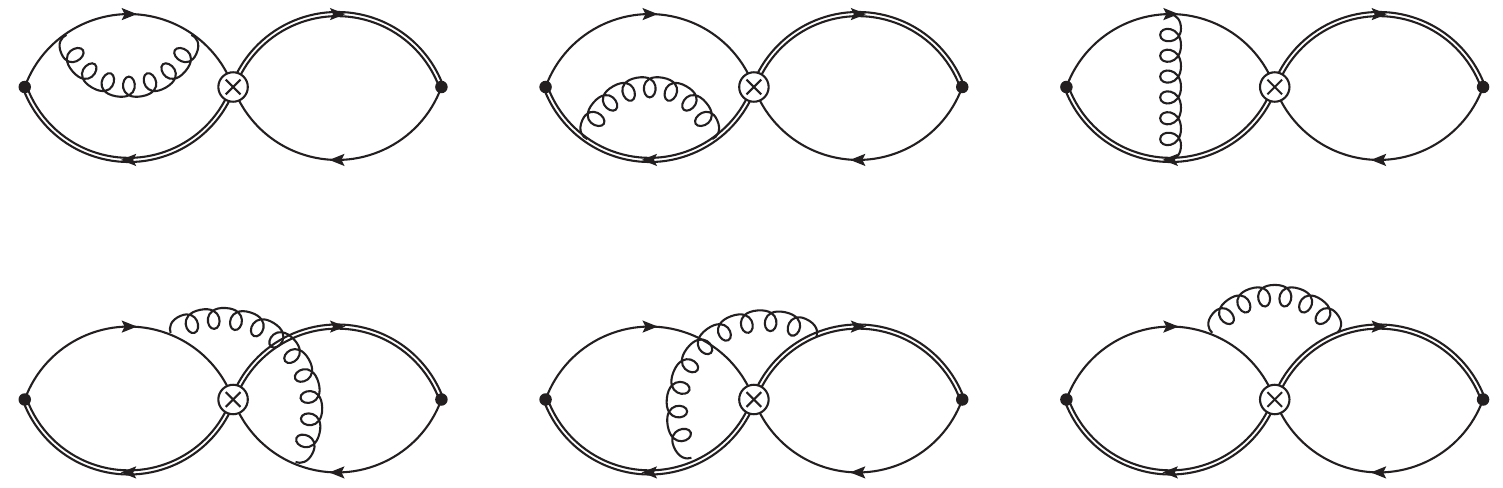}
  \caption{\label{fig:ThreePointCorrelator}
  Diagrams contributing to the three-point HQET correlator~\eqref{eq:DefK} at NLO. 
  Symmetric diagrams are not shown.}
 \end{center}
\end{figure}

The bare NLO correction $K_{\tilde{Q}}^{(1),\text{bare}}$ is given by the 
diagrams shown in Figure~\ref{fig:ThreePointCorrelator}. At this order we 
get corrections that do not factorize due to gluon exchange between the 
left and right-hand side. These genuine three-loop contributions -- given 
by the diagrams in the second row of Figure~\ref{fig:ThreePointCorrelator} 
-- are the most computationally challenging. The Dirac traces have been 
evaluated with both \texttt{TRACER}~\cite{Jamin:1991dp} and 
\texttt{Package-X}~\cite{Patel:2015tea,Patel:2016fam}. 
We use the code \texttt{FIRE}~\cite{Smirnov:2008iw,Smirnov:2013dia,Smirnov:2014hma} 
to find IBP relations~\cite{Chetyrkin:1981qh} between the three-loop integrals 
and to reduce them to a set of master integrals via the Laporta 
algorithm~\cite{Laporta:2001dd}. The relevant master integrals have been 
computed analytically in~\cite{Grozin:2008nu,Grozin:2016uqy}. 

The renormalized NLO correlators are given by 
\begin{equation}
 K_{\tilde{Q}_i}^{(1)} = K_{\tilde{Q}_i}^{(1),\text{bare}} + \frac{1}{2\epsilon}\left[
                           \left(2\tilde{\gamma}_{\tilde{j}}^{(0)}\delta_{ij}+\tilde{\gamma}_{\tilde{Q}_i\tilde{Q}_j}^{(0)}\right)K_{\tilde{Q}_j}^{(0)}
                           +\tilde{\gamma}_{\tilde{Q}_i\tilde{E}_j}^{(0)}K_{\tilde{E}_j}^{(0)}\right],
\end{equation}
where $\tilde{\gamma}_{\tilde{j}}^{(0)}=-3C_F$ is the LO anomalous dimension of 
the currents $\tilde{j}_\pm$. The contributions from the evanescent operators 
modify the double discontinuities of the correlators by a finite amount and 
introduce a dependence of the correlator on the choice of basis of the HQET 
evanescent operators. This dependence propagates to the HQET bag parameters 
extracted in the sum rule and cancels with the HQET evanescent scheme dependence 
of the matching coefficients~\eqref{eq:CQQ}  
in the matching equation~\eqref{eq:B_QCD_HQET_matching} for the QCD Bag parameters. 
The results for the bare correlators are available as an ancillary \texttt{Mathematica} 
file with the arXiv version of this article. Here, we only show the compact results 
for the double discontinuities of the correlators. 

Methods to compute the double discontinuities of the correlators have been 
described in~\cite{Ball:1993xv,Grozin:2016uqy}. The results take the form 
\begin{equation}
 \rho_{\tilde{Q}_i}^\text{pert}(\omega_1,\omega_2) = A_{\tilde{Q}_i}\rho_\Pi(\omega_1)\rho_\Pi(\omega_2)+\Delta\rho_{\tilde{Q}_i},
 \label{eq:rho_decomposition}
\end{equation}
where 
\begin{eqnarray}
 \rho_\Pi(\omega) & \equiv & \frac{\Pi(\omega+i0)-\Pi(\omega-i0)}{2\pi i} \nonumber\\
                  &    =   & \frac{N_c\omega^2}{2\pi^2}\left[1+\frac{\alpha_sC_F}{4\pi}\left(17+\frac{4\pi^2}{3}+3\ln\frac{\mu^2}{4\omega^2}\right)+\mathcal{O}(\alpha_s^2)\right],
\end{eqnarray}
is the discontinuity of the two-point correlator~\eqref{eq:Pi} up to two-loop 
order~\cite{Broadhurst:1991fc,Bagan:1991sg,Neubert:1991sp}. 
The non-factorizable contributions are 
\begin{equation}
 \Delta\rho_{\tilde{Q}_i} \equiv \frac{N_c C_F}{4} \frac{\omega_1^2\omega_2^2}{\pi^4}\frac{\alpha_s}{4\pi}r_{\tilde{Q}_i}(x,L_\omega),
 \label{eq:r_def}
\end{equation}
where $x=\omega_2/\omega_1$, $L_\omega=\ln(\mu^2/(4\omega_1\omega_2))$ and we obtain 
\begin{eqnarray}
 r_{\tilde{Q}_1}(x,L_\omega) & = & 8-\frac{a_2}{2}-\frac{8 \pi ^2}{3},\nonumber\\
 r_{\tilde{Q}_2}(x,L_\omega) & = & 25+\frac{a_1}{2}-\frac{4 \pi ^2}{3}+6 L_\omega+\phi(x),\nonumber\\
 r_{\tilde{Q}_4}(x,L_\omega) & = & 16-\frac{a_3}{4}-\frac{4 \pi ^2}{3}+3L_\omega+\frac{\phi(x)}{2},\nonumber\\
 r_{\tilde{Q}_5}(x,L_\omega) & = & 29-\frac{a_3}{2}-\frac{8 \pi ^2}{3}+6 L_\omega+\phi(x),
\end{eqnarray}
where 
\begin{equation}
 \phi(x)=\begin{cases}
  x^2-8 x+6 \ln(x),\hspace{1cm}x\leq1,\\ 
  \frac{1}{x^2}-\frac{8}{x}-6 \ln(x),\hspace{1.2cm}x>1.
 \end{cases}
\end{equation}
Taking $a_2=-4$ in accordance with \cite{Grozin:2016uqy} we reproduce their 
result for $r_{\tilde{Q}_1}$ up to a factor of 2 which is due to the different 
normalization of the HQET operators.

\subsection{Sum rule for the Bag parameters\label{sec:sum_rule_bags}}

Inserting the decomposition~\eqref{eq:rho_decomposition} into the sum 
rule~\eqref{eq:SR_matrix_elements} allows us to subtract the factorized 
contribution using the sum rule~\cite{Broadhurst:1991fc,Bagan:1991sg,Neubert:1991sp} 
for the HQET decay constant 
\begin{equation}
 F^2(\mu)e^{-\frac{\Lbar}{t}}=\int\limits_0^{\omega_c}d\omega e^{-\frac{\omega}{t}}\rho_\Pi(\omega) + \dots \, . 
 \label{eq:SR_F2}
\end{equation}
The factorizable part of~\eqref{eq:rho_decomposition} exactly reproduces 
the VSA for the matrix elements. After subtracting it, we obtain a sum rule 
for the deviation $\Delta B_{\tilde{Q}} = B_{\tilde{Q}}-1$ from the VSA. 
In the traditional sum rule approach this gives 
\begin{eqnarray}
\label{eq:B_OneInt}
\Delta B_{\tilde{Q}_i} & = & \frac{1}{A_{\tilde{Q}_i}F(\mu)^4}
\int\limits_0^{\omega_c}d\omega_1d\omega_2 e^{\frac{\Lbar-\omega_1}{t_1}+\frac{\Lbar-\omega_2}{t_2}}\Delta\rho_{\tilde{Q}_i}(\omega_1,\omega_2)\\
& = & \frac{1}{A_{\tilde{Q}_i}}\frac{\int\limits_0^{\omega_c}d\omega_1d\omega_2 e^{-\frac{\omega_1}{t_1}-\frac{\omega_2}{t_2}}\Delta\rho_{\tilde{Q}_i}(\omega_1,\omega_2)
      }{\left(\int\limits_0^{\omega_c}d\omega_1e^{-\frac{\omega_1}{t_1}}\rho_\Pi(\omega_1)\right)\left(\int\limits_0^{\omega_c}d\omega_2e^{-\frac{\omega_2}{t_2}}\rho_\Pi(\omega_2)\right)}.
\label{eq:B_ThreeInt}
\end{eqnarray}
The stability of the sum rule~\eqref{eq:B_ThreeInt} can then be assessed 
numerically by variation of the cutoff $\omega_c$ and the Borel parameters 
$t_i$, see e.g.~\cite{Bagan:1991sg,Ball:1993xv}. 

In our analysis we follow a different 
approach that allows us to obtain analytic results for the HQET Bag parameters. 
This exploits the fact that the dispersion relation~\eqref{eq:disp_rel} 
is not violated by the introduction of an arbitrary weight function 
$w(\omega_1,\omega_2)$ in the integration as long as it is chosen such that 
no additional discontinuities appear in the complex plane.\footnote{
The arbitrariness of the weight function is a mathematical statement 
which holds for the dispersion relation. The sum rule~\eqref{eq:sum_rule_uncut} 
does however also assume quark-hadron duality and breaks down if 
pathological weight functions are used, e.g. rapidly oscillating ones. 
In the following we only use slowly varying weight functions with support 
on the complete integration domain.} 
In the presence of such a weight function $w$ the square of the 
sum rule~\eqref{eq:SR_F2} takes the form  
\begin{equation}
 F^4(\mu)e^{-\frac{\Lbar}{t_1}-\frac{\Lbar}{t_2}}w(\Lbar,\Lbar) = 
    \int\limits_0^{\omega_c}d\omega_1d\omega_2 e^{-\frac{\omega_1}{t_1}-\frac{\omega_2}{t_2}} 
    w(\omega_1,\omega_2)\rho_\Pi(\omega_1)\rho_\Pi(\omega_2) + \dots \, . 
 \label{eq:SR_F4_weighted}
\end{equation}
Since the condensate contributions have already been taken into account 
in~\cite{Grozin:2016uqy,Mannel:2007am,Mannel:2011zza} and are in the 
subpercent range we only focus on the perturbative contribution to 
the sum rule. By using~\eqref{eq:SR_F4_weighted} with the choice 
\begin{equation}
 w_{\tilde{Q}_i}(\omega_1,\omega_2) = 
    \frac{\Delta\rho_{\tilde{Q}_i}^\text{pert}(\omega_1,\omega_2)}{\rho_\Pi^\text{pert}(\omega_1)\rho_\Pi^\text{pert}(\omega_2)} = 
    \frac{C_F}{N_c} \, \frac{\alpha_s}{4\pi} \, r_{\tilde{Q}_i}(x,L_\omega),
 \label{eq:weight_function}
\end{equation}
we can remove the integration in~\eqref{eq:B_OneInt} altogether 
and find the simple result 
\begin{equation}
 \Delta B_{\tilde{Q}_i}^\text{pert}(\mu_\rho) = \frac{C_F}{N_c A_{\tilde{Q}_i}} \, \frac{\alpha_s(\mu_\rho)}{4\pi} \, r_{\tilde{Q}_i}\left(1,\log\frac{\mu_\rho^2}{4\Lbar^2}\right). 
 \label{eq:B_NoInt}
\end{equation}

The sum rule is valid at a low scale $\mu_\rho\sim 2\omega_i\sim2\Lbar$ 
where the logarithms that appear in the spectral functions are small. 
From there we have to evolve the results for the Bag parameters up to 
the scale $\mu_m\sim m_b$ where the matching~\eqref{eq:B_QCD_HQET_matching} 
to the QCD Bag parameters can be performed without introducing large 
logarithms. From~\eqref{eq:Bags_HQET} and the running of the HQET operators 
and decay constant 
\begin{equation}
 \frac{d\vec{\tilde{Q}}}{d\,\ln\mu} = -\hat{\tilde{\gamma}}_{\tilde{Q}\tilde{Q}}\,\vec{\tilde{Q}}, \hspace{1cm}\frac{d F(\mu)}{d\,\ln\mu} = -\tilde{\gamma}_{\tilde{j}}F(\mu),
\end{equation}
we obtain the RG equations for the HQET Bag parameters 
\begin{equation}
 \frac{d\vec{B}_{\tilde{Q}}}{d\,\ln\mu} = -\left(\hat{A}_{\tilde{Q}}^{-1}\hat{\tilde{\gamma}}_{\tilde{Q}\tilde{Q}}\hat{A}_{\tilde{Q}}-2\tilde{\gamma}_{\tilde{j}}\right)\vec{B}_{\tilde{Q}} 
 \equiv -\hat{\tilde{\gamma}}_{\tilde{B}}\vec{B}_{\tilde{Q}}, 
 \label{eq:RGE_HQET_Bags}
\end{equation}
where $\hat{A}_{\tilde{Q}}$ is the diagonal matrix with 
entries $A_{\tilde{Q}}$ given in~\eqref{eq:AQi_HQET}. 
The LO solution to~\eqref{eq:RGE_HQET_Bags} takes the form 
\begin{equation}
 \vec{B}_{\tilde{Q}}(\mu) = \hat{U}_{\tilde{B}}^{(0)}(\mu,\mu_0)\,\vec{B}_{\tilde{Q}}(\mu_0),
 \label{eq:Running_HQET_Bags}
\end{equation}
with the LO evolution matrix
\begin{equation}
 \hat{U}_{\tilde{B}}^{(0)}(\mu,\mu_0) = \left(\frac{\alpha_s(\mu)}{\alpha_s(\mu_0)}\right)^{\frac{\hat{\tilde{\gamma}}_{\tilde{B}}^{(0)}}{2\beta_0}} 
 = \hat{V}\,\left(\frac{\alpha_s(\mu)}{\alpha_s(\mu_0)}\right)^{\frac{\vec{\tilde{\gamma}}_{\tilde{B}}^{(0)}}{2\beta_0}} \,\hat{V}^{-1}, 
 \label{eq:U_HQET_Bags}
\end{equation}
where $\hat{V}$ is the transformation that diagonalizes the ADM 
$\hat{\tilde{\gamma}}_{\tilde{B}}^{(0)}$ 
\begin{equation}
 \hat{\tilde{\gamma}}_{\tilde{B}}^{(0),\text{D}} = \hat{V}^{-1}\hat{\tilde{\gamma}}_{\tilde{B}}^{(0)}\hat{V},
\end{equation}
and the vector $\vec{\tilde{\gamma}}_{\tilde{B}}^{(0)}$ contains the diagonal 
entries of $\hat{\tilde{\gamma}}_{\tilde{B}}^{(0),\text{D}}$. 
As part of our error analysis we allow the matching scale $\mu_m$ to differ 
from $\overline{m}_b(\overline{m}_b)$ and then evolve the QCD Bag parameters back to 
$\overline{m}_b(\overline{m}_b)$. The LO evolution matrix has the same form as its HQET 
counterpart~\eqref{eq:U_HQET_Bags} while the anomalous dimension matrix 
of the QCD Bag parameters is given by 
\begin{equation}
 \hat{\gamma}_{B} = \hat{A}_{Q}^{-1}\hat{\gamma}_{QQ}\hat{A}_{Q}. 
\end{equation}
We only resum the leading logarithms because the NLO anomalous dimensions in HQET 
are currently not known. This implies that dependence of the QCD matrix elements 
on the basis of evanescent HQET operators does not fully cancel. As discussed below, 
we use variation of the parameters $a_i$ to estimate the effects of NLL resummation. 
We expect this effect to be small since the scales $\mu_\rho$ and $\mu_m$ are not 
very widely separated and $\ln(\mu_m/\mu_\rho)$ is of order one.


\section{\boldmath Results for $\Delta B=2$ operators\label{sec:results}}

We describe our analysis in Section~\ref{sec:results_details} and give 
the results for the Bag parameters, together with a comparison with 
other works, in Section~\ref{sec:results_results}. 
In Section~\ref{sec:results_mixing} the results for the mixing 
observables with our Bag parameters are shown. 

\subsection{Details of the analysis\label{sec:results_details}}

We determine the HQET Bag parameters from the sum rule~\eqref{eq:B_NoInt} with 
the central values $\mu_\rho = 1.5\text{ GeV}$ and $\Lbar = 0.5\text{ GeV}$. 
We use \texttt{RunDec}~\cite{Chetyrkin:2000yt,Herren:2017osy} to evolve 
$\alpha_s(M_Z)=0.1181$~\cite{Olive:2016xmw} down to the bottom-quark 
$\overline{\text{MS}}$ mass 
$\overline{m}_b(\overline{m}_b) = 4.203\text{ GeV}$~\cite{Beneke:2014pta,Beneke:2016oox} 
with five-loop accuracy~\cite{Baikov:2016tgj,Herzog:2017ohr,Luthe:2017ttc,Luthe:2017ttg,Chetyrkin:2017bjc}. 
From there we use two-loop running with four and five flavours in HQET and QCD, 
respectively. The decoupling of the bottom quark is trivial at this accuracy. 

The HQET Bag parameters are then evolved from the scale $\mu_\rho$ up to the 
scale $\mu_m = \overline{m}_b(\overline{m}_b)$ using~\eqref{eq:Running_HQET_Bags}. 
There the matching~\eqref{eq:B_QCD_HQET_matching} to the QCD Bag parameters 
is performed. The factors $C_{Q_i\tilde{Q}_j}(\mu)/C^2(\mu)$ are expanded in 
$\alpha_s$ and truncated after the linear term. We also expand the ratios 
$A_{\tilde{Q}_j}/A_{Q_i}$ strictly in $\Lbar/m_b$ and $m_q/m_b$. 
Up to higher order perturbative corrections, this is equivalent to the 
use of the VSA for the power-suppressed HQET operators that arise in the 
QCD-HQET matching~\eqref{eq:matching_condition}. 

A small dependence on the choice of basis for the evanescent HQET operators 
remains in the QCD Bag parameters because the RG evolution of the HQET Bag 
parameters is only known at the LL level. We have checked that the 
$a_i$-dependence fully cancels when the scales $\mu_\rho$ and $\mu_m$ are 
identified and the matching~\eqref{eq:B_QCD_HQET_matching} is strictly 
expanded in the strong coupling, which serves as a strong cross-check of 
our calculation. For different scales $\mu_\rho$ and $\mu_m$ the remaining 
$a_i$-dependence can be removed by a future computation of the NLO ADMs. 

Finally, we convert the QCD Bag parameters $B_{Q}$ to the usual 
convention $\overline{B}_Q$ defined in~\eqref{eq:Bags_QCD_MSmasses}. 
This is done by expanding the ratios of the prefactors 
$A_Q/\overline{A}_Q(\overline{m}_b(\overline{m}_b))$ in $\alpha_s$ and 
truncating them after the linear term. 

To estimate the errors of the Bag parameters we take the following 
sources of uncertainties into account: 
\begin{itemize}
 \item The uncertainty in the analytic form~\eqref{eq:B_NoInt} of the sum 
 rule is estimated through variation of the residual mass $\Lbar$ in the 
 range [0.4,0.6] GeV. 
 In addition we include an intrinsic sum rule uncertainty of $0.02$ in the HQET 
 bag parameters. The numerical value is determined from the comparison 
 of the analytic values~\eqref{eq:B_NoInt} with results obtained from 
 the traditional sum rule approach~\eqref{eq:B_ThreeInt}. 
 \item The condensate contributions to $B_{\tilde{Q}_1}$ and $B_{\tilde{Q}_2}$ 
 are taken from~\cite{Mannel:2007am,Mannel:2011zza} and are in the subpercent 
 range. For $B_{\tilde{Q}_4}$ and $B_{\tilde{Q}_5}$, which have not been 
 determined there, we therefore add an error of $\pm0.01$ to the perturbative 
 results. 
 \item To assign an uncertainty from the unknown $\alpha_s^2$ contributions 
 to the spectral densities we vary the scale $\mu_\rho$ in the range 
 [1,2] GeV. 
 \item As discussed above we implicitly include higher-order corrections 
 in $1/m_b$ in the VSA approximation. The non-factorizable corrections 
 of this kind are of the order $(\alpha_s/\pi)\cdot(\Lbar/m_b) \sim 0.01$, 
 which we take as an estimate for the error. 
 \item Higher order perturbative contributions to the QCD-HQET matching 
 relation and the RG evolution of the Bag parameters are estimated 
 through variation of $\mu_m$ in the range [3,6] GeV and variation 
 of the $a_i$ in the range [-10,10]. The QCD Bag parameters are then 
 evolved to the central scale $\overline{m}_b(\overline{m}_b)$ with LL 
 accuracy as described in Section~\ref{sec:sum_rule_bags}. 
 
 The variation of $\mu_m$ by the usual factors of 1/2 and 2 would lead to 
 a doubling of the matching uncertainty estimates given below, which would 
 significantly exceed the effect of the NLO matching at the central scale. 
 We therefore use a less conservative range but cannot exclude larger matching 
 effects at NNLO at present, while a calculation is not available. 
 \item The parametric uncertainty from $\alpha_s(M_Z)$ is in the permille 
 range and neglected. 
\end{itemize}
The individual errors are then summed in quadrature. We also divide the 
uncertainties into a sum rule uncertainty which contains the first three 
items in the list above and a matching uncertainty which contains the 
remaining three.

\subsection{Results and comparison\label{sec:results_results}}

From the sum rule we obtain the HQET Bag parameters 
\begin{equation}
\begin{aligned}
 B_{\tilde{Q}_1} (1.5\text{ GeV}) & = 0.910\,_{-0.031}^{+0.023} = 0.910\,_{-0.000}^{+0.000}(\Lbar)\,_{-0.020}^{+0.020}(\text{intr.})\,_{-0.005}^{+0.005}(\text{cond.})\,_{-0.024}^{+0.011}(\mu_\rho),\vspace*{0.1cm}\\
 B_{\tilde{Q}_2} (1.5\text{ GeV}) & = 0.923\,_{-0.035}^{+0.029} = 0.923\,_{-0.020}^{+0.016}(\Lbar)\,_{-0.020}^{+0.020}(\text{intr.})\,_{-0.004}^{+0.004}(\text{cond.})\,_{-0.020}^{+0.013}(\mu_\rho),\vspace*{0.1cm}\\
 B_{\tilde{Q}_4} (1.5\text{ GeV}) & = 1.009\,_{-0.023}^{+0.024} = 1.009\,_{-0.006}^{+0.007}(\Lbar)\,_{-0.020}^{+0.020}(\text{intr.})\,_{-0.010}^{+0.010}(\text{cond.})\,_{-0.003}^{+0.003}(\mu_\rho),\vspace*{0.1cm}\\
 B_{\tilde{Q}_5} (1.5\text{ GeV}) & = 1.004\,_{-0.028}^{+0.030} = 1.004\,_{-0.016}^{+0.020}(\Lbar)\,_{-0.020}^{+0.020}(\text{intr.})\,_{-0.010}^{+0.010}(\text{cond.})\,_{-0.006}^{+0.004}(\mu_\rho),\vspace*{0.3cm}
 \label{eq:DelB2_HQET_results}
\end{aligned}
\end{equation}
where we have set $a_i=0$ for $i=1,2,3$ to specify a unique basis of evanescent 
HQET operators. The individual uncertainties were determined as described above 
and added in quadrature. The corrections to the VSA for scales in the range from 
1\,-\,2 GeV are at the level of 5\,-\,11\,\% for $\tilde{Q}_{1,2}$ and 0\,-\,4\,\% 
for $\tilde{Q}_{4,5}$. We find that the total sum rule uncertainties of the 
Bag parameters are quite small. This is because the sum rule~\eqref{eq:B_NoInt} 
is formulated for the deviation from the VSA and the substantial relative 
uncertainties of the sum rule itself are small in comparison with the VSA 
contribution to the Bag parameters. 

Following the steps outlined in Section~\ref{sec:results_details} 
we obtain the following results for the QCD Bag parameters 
\begin{eqnarray}
 \overline{B}_{Q_1} (\overline{m}_b(\overline{m}_b)) & = & 0.868\,_{-0.050}^{+0.051} = 0.868\,_{-0.029}^{+0.021}(\text{sum rule})\,_{-0.041}^{+0.046}(\text{matching}),\vspace*{0.1cm}\nonumber\\
 \overline{B}_{Q_2} (\overline{m}_b(\overline{m}_b)) & = & 0.842\,_{-0.073}^{+0.078} = 0.842\,_{-0.033}^{+0.028}(\text{sum rule})\,_{-0.065}^{+0.073}(\text{matching}),\vspace*{0.1cm}\nonumber\\
 \overline{B}_{Q_3} (\overline{m}_b(\overline{m}_b)) & = & 0.818\,_{-0.159}^{+0.162} = 0.818\,_{-0.132}^{+0.126}(\text{sum rule})\,_{-0.087}^{+0.102}(\text{matching}),\vspace*{0.1cm}\nonumber\\
 \overline{B}_{Q_4} (\overline{m}_b(\overline{m}_b)) & = & 1.049\,_{-0.084}^{+0.092} = 1.049\,_{-0.025}^{+0.025}(\text{sum rule})\,_{-0.080}^{+0.089}(\text{matching}),\vspace*{0.1cm}\nonumber\\
 \overline{B}_{Q_5} (\overline{m}_b(\overline{m}_b)) & = & 1.073\,_{-0.075}^{+0.083} = 1.073\,_{-0.026}^{+0.028}(\text{sum rule})\,_{-0.070}^{+0.078}(\text{matching}).
 \label{eq:DelB2_results}
\end{eqnarray}
The evolution to the scale $\overline{m}_b(\overline{m}_b))$ and the matching 
to QCD increase the deviations from the VSA to up to $18\,\%$. With the exception 
of $\overline{B}_{Q_3}$ the uncertainties of the Bag parameters are dominated 
by the matching. A detailed list of the uncertainties can be found in 
Appendix~\ref{sec:Uncertainties}. 

In Figure~\ref{fig:DelB2_comparison} we compare our results to other 
recent determinations from lattice simulations~\cite{Dalgic:2006gp,
Carrasco:2013zta,Bazavov:2016nty,Aoki:2016frl} and sum rules~\cite{Grozin:2016uqy}. 
\begin{figure}[t]
 \begin{center}
    \includegraphics[width=\textwidth]{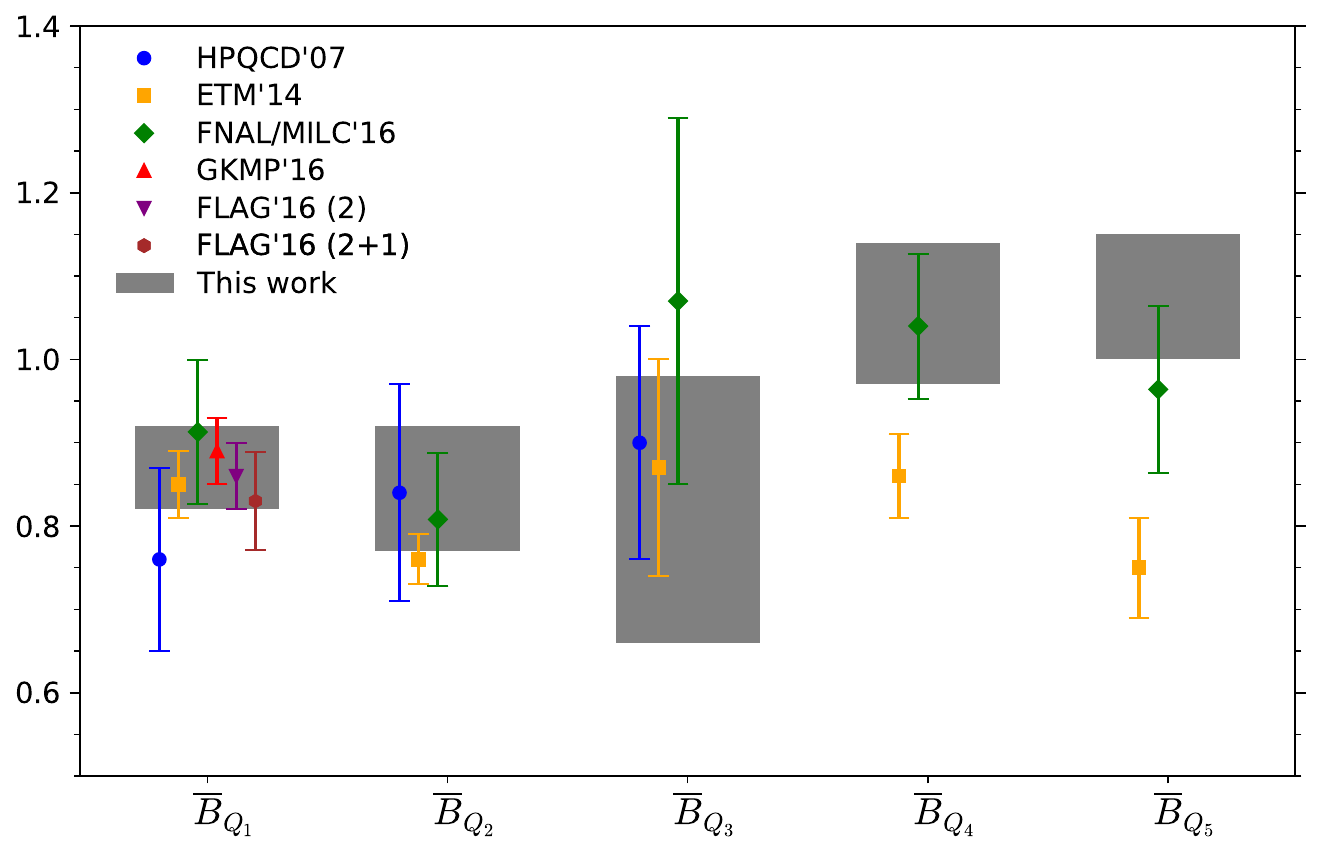}
  \caption{Comparison of our results for the $\Delta B=2$ Bag parameters at the 
  scale $\overline{m}_b(\overline{m}_b)$ to the lattice values of HPQCD'07~\cite{Dalgic:2006gp}, 
  ETM'14~\cite{Carrasco:2013zta} and FNAL/MILC'16~\cite{Bazavov:2016nty}, 
  the FLAG averages~\cite{Aoki:2016frl} and the sum rule result 
  GKMP'16~\cite{Grozin:2016uqy}. 
  }
  \label{fig:DelB2_comparison}
 \end{center}
\end{figure}
We find excellent agreement for the Bag parameters of the operators 
$Q_1$, $Q_2$ and $Q_3$. The uncertainties of our sum rule analysis 
are similar to those obtained on the lattice. We observe that the 
uncertainty of the Bag parameter $\overline{B}_{Q_3}$ is significantly 
larger than those of $\overline{B}_{Q_1}$ and $\overline{B}_{Q_2}$. 
This is related to the small color factor $A_{Q_3}=1/3 + \mathcal{O}(1/m_b)$ 
which implies that the sum rule uncertainties get enhanced by the 
factors $A_{\tilde{Q}_1}/A_{Q_3} = 8 + \mathcal{O}(1/m_b)$ and 
$A_{\tilde{Q}_2}/A_{Q_3} = -5 + \mathcal{O}(1/m_b)$ in the 
matching~\eqref{eq:B_QCD_HQET_matching} of the Bag parameters. 
The absolute sum rule uncertainty of the matrix element of $Q_3$ 
is of a similar size as that of the other operators. 

The tiny difference of the central value of $\overline{B}_{Q_1}$ 
compared to the sum rule determination~\cite{Grozin:2016uqy} is mostly 
due to different scale choices. Since $\overline{B}_{Q_1}$ does not 
run at the LL order, \cite{Grozin:2016uqy} sets all scales equal to 
the bottom-quark mass. We, however, evaluate the sum rule at a lower 
scale $\mu_\rho\sim1.5\text{ GeV}$ where the strong coupling is larger 
and causes a bigger deviation from the VSA. 

Only two previous lattice results~\cite{Carrasco:2013zta,Bazavov:2016nty} 
exist for the matrix elements of the operators $Q_4$ and $Q_5$, and they 
differ at the level of more than two sigma. Our results are in very good 
agreement with those of~\cite{Bazavov:2016nty} and show an even higher 
level of tension with~\cite{Carrasco:2013zta} in $\overline{B}_{Q_5}$.

\subsection{\boldmath$B_s$ and $B_d$ mixing observables\label{sec:results_mixing}}

We consider the mass and decay rate differences $\Delta M_s = M_H^s - M_L^s$ 
and $\Delta\Gamma_s = \Gamma_L^s - \Gamma_H^s$, where $M_{H/L}^s$ and 
$\Gamma_{H/L}^s$ are the mass and width of the heavy (H) and light (L) 
physical eigenstates of the $B_s$\,-\,$\bar{B}_s$ system, as well as the 
semileptonic decay asymmetry 
\begin{equation}
 a_\text{sl}^{s} = \frac{\Gamma(\bar{B}_s(t)\to f)-\Gamma(B_s(t)\to \bar{f})}{\Gamma(\bar{B}_s(t)\to f)+\Gamma(B_s(t)\to \bar{f})},
\end{equation}
where $f$ is a flavor-specific final state, i.e. $\bar{B}_s\to f$ and 
$B_s\to\bar{f}$ are forbidden (see~\cite{Artuso:2015swg} for a recent 
review of $B_s$ mixing). Using our values for the Bag parameters, we 
give predictions for these observables and compare them to the current 
experimental averages given by the HFLAV~\cite{Amhis:2016xyh}. In our 
sum rule determination we have assumed the light quark $q$ in the $B_q$ 
meson to be massless. The corrections to~\eqref{eq:B_NoInt} from a 
non-zero strange-quark mass are of the order $(\alpha_s/\pi)(m_s/(2\Lbar))\approx0.02$. 
This point has recently been discussed in more detail in~\cite{Grozin:2017uto}. 
We add another uncertainty of $\pm0.02$ in quadrature to the 
results~\eqref{eq:DelB2_results} to account for the unknown corrections. 
The effect on the total uncertainty is small. 

We find good agreement between experiment and the SM prediction for the 
mass difference: 
\begin{equation}
 \begin{array}{lll}
 \Delta M_s^\text{exp} & = (17.757 \pm 0.021)\,\text{ps}^{-1},\hspace{-2.5cm} & \vspace*{0.1cm}\\
 \Delta M_s^\text{SM}  & = (16.6\pm1.7)\,\text{ps}^{-1} & = (16.6_{-1.1}^{+1.2}\,\,(\text{had.})\pm 0.1\,(\text{scale})_{-1.2}^{+1.3}\,\,(\text{param.}))\,\text{ps}^{-1},
 \end{array}
\end{equation}
where we have used the input values given in Appendix~\ref{sec:Uncertainties}. 
The 10\% uncertainty of the SM prediction is dominated by the hadronic 
and parametric CKM uncertainties which are of the same size. 
We also give results for the mass difference in the $B_d$ system 
\begin{equation}
 \begin{array}{lll}
 \Delta M_d^\text{exp} & = (0.5065 \pm 0.0019)\,\text{ps}^{-1},\hspace{-2.5cm} & \vspace*{0.1cm}\\
 \Delta M_d^\text{SM}  & = (0.56\pm 0.08)\,\text{ps}^{-1} & = (0.56\pm 0.04\,(\text{had.})\pm 0.00\,(\text{scale})\pm 0.07\,(\text{param.}))\,\text{ps}^{-1}, 
 \end{array}
\end{equation}
which is in good agreement with the experimental value. 

We determine the decay rate difference and the semileptonic decay 
asymmetry in the $\overline{\text{MS}}$, PS~\cite{Beneke:1998rk}, 
1S~\cite{Hoang:1998ng} and kinetic~\cite{Bigi:1996si} mass schemes 
with the mass values given in Appendix~\ref{sec:Uncertainties}. 
The $\overline{\text{MS}}$ charm-quark mass at the scale of the bottom-quark 
mass has been used throughout. We obtain 
\begin{equation}
 \begin{array}{lll}
 \Delta \Gamma_s^\text{exp}             & = (0.090 \pm 0.005)\,\text{ps}^{-1},\hspace{-1.5cm} & \vspace*{0.1cm}\\
 \Delta \Gamma_s^{\overline{\text{MS}}} & = (0.080_{-0.023}^{+0.018})\,\text{ps}^{-1} & = (0.080\pm 0.016\,(\text{had.})_{-0.015}^{+0.006}\,\,(\text{scale})\pm 0.006\,(\text{param.}))\,\text{ps}^{-1}, \vspace{0.1cm}\\
 \Delta \Gamma_s^\text{PS}              & = (0.079_{-0.026}^{+0.020})\,\text{ps}^{-1} & = (0.079\pm 0.018\,(\text{had.})_{-0.018}^{+0.007}\,\,(\text{scale})\pm 0.006\,(\text{param.}))\,\text{ps}^{-1}, \vspace{0.1cm}\\
 \Delta \Gamma_s^\text{1S}              & = (0.075_{-0.028}^{+0.021})\,\text{ps}^{-1} & = (0.075\pm 0.019\,(\text{had.})_{-0.020}^{+0.008}\,\,(\text{scale})\pm 0.006\,(\text{param.}))\,\text{ps}^{-1}, \vspace{0.1cm}\\
 \Delta \Gamma_s^\text{kin}             & = (0.076_{-0.027}^{+0.020})\,\text{ps}^{-1} & = (0.076\pm 0.018\,(\text{had.})_{-0.019}^{+0.008}\,\,(\text{scale})\pm 0.006\,(\text{param.}))\,\text{ps}^{-1},
 \end{array}
\end{equation}
and 
\begin{equation}
 \begin{array}{lll}
 a_\text{sl}^{s,\,\text{exp}}           & = (-60 \pm 280)\cdot10^{-5},\hspace{-1.5cm} & \vspace*{0.1cm}\\
 a_\text{sl}^{s,\,\overline{\text{MS}}} & = (2.3\pm 0.3)\cdot10^{-5} & = (2.3\pm 0.1\,(\text{had.})_{-0.1}^{+0.0}\,\,(\text{scale})\pm 0.3\,(\text{param.}))\cdot10^{-5}, \vspace{0.1cm}\\
 a_\text{sl}^{s,\,\text{PS}}            & = (2.2\pm 0.3)\cdot10^{-5} & = (2.2\pm 0.1\,(\text{had.})_{-0.1}^{+0.0}\,\,(\text{scale})_{-0.3}^{+0.2}\,\,(\text{param.}))\cdot10^{-5}, \vspace{0.1cm}\\
 a_\text{sl}^{s,\,\text{1S}}            & = (2.1_{-0.3}^{+0.2})\cdot10^{-5} & = (2.1\pm 0.1\,(\text{had.})_{-0.1}^{+0.0}\,\,(\text{scale})_{-0.3}^{+0.2}\,\,(\text{param.}))\cdot10^{-5}, \vspace{0.1cm}\\
 a_\text{sl}^{s,\,\text{kin}}           & = (2.2\pm0.3)\cdot10^{-5} & = (2.2\pm 0.1\,(\text{had.})_{-0.1}^{+0.0}\,\,(\text{scale})_{-0.3}^{+0.2}\,\,(\text{param.}))\cdot10^{-5}.
 \end{array}
\end{equation}
The different mass schemes are in good agreement with each other 
and we adopt the PS mass scheme as our central result. The SM value 
for the decay rate difference is in good agreement with the experimental 
average. The theory uncertainty is currently at the level of 30\%. 
It is dominated by the matrix elements of the dimension seven operators, 
in particular the VSA estimate $\overline{B}_{R_2}=1\pm0.5$ contributes 
$\pm0.016\,\text{ps}^{-1}$ to the uncertainty. The second largest 
contribution is the scale variation. A detailed overview is given in 
Appendix~\ref{sec:Uncertainties}. To achieve a significant reduction 
of the combined uncertainties, a determination of the dimension seven 
matrix elements and a NNLO calculation of the perturbative matching are 
needed. 

The experimental uncertainty for the semileptonic decay asymmetry 
is two orders of magnitude larger than the SM prediction, which makes 
this a clear null test for the SM~\cite{Laplace:2002ik}. The decay 
rate difference and the semileptonic decay asymmetry in the $B_d$ 
system have also not been measured yet. The current experimental 
averages and our predictions are 
\begin{equation}
 \begin{array}{ll}
 \Delta \Gamma_d^\text{exp} & = (-1.3\pm6.6)\cdot10^{-3}\,\text{ps}^{-1}, \vspace{0.1cm}\\
 \Delta \Gamma_d^\text{PS} & = (2.7_{-0.9}^{+0.7})\cdot10^{-3}\,\text{ps}^{-1} = (2.7_{-0.6}^{+0.6}\,\,(\text{had.})_{-0.6}^{+0.2}\,\,(\text{scale})_{-0.4}^{+0.4}\,\,(\text{param.}))\cdot10^{-3}\,\text{ps}^{-1},\vspace{0.3cm}\\
 a_\text{sl}^{d,\,\text{exp}} & = (-21\pm 17)\cdot10^{-4}, \vspace{0.1cm}\\
 a_\text{sl}^{d,\,\text{PS}} & = (-4.4_{-0.5}^{+0.6})\cdot10^{-4} = (-4.4\pm 0.1\,(\text{had.})_{-0.1}^{+0.2}\,\,(\text{scale})\pm 0.5\,(\text{param.}))\cdot10^{-4}. 
 \end{array}
\end{equation}
The results obtained in different mass schemes are compatible 
and the relative uncertainties of the predictions are of the same 
magnitude as in the $B_s$ system.


\section{\boldmath$\Delta B=0$ operators and ratios of $B$-meson lifetimes\label{sec:lifetimes}}

The dominant contribution to lifetime differences between the mesons 
$B_q$ with $q=u,d,s$ is due to spectator effects which first appear 
as dimension-six contributions in the HQE. The NLO Wilson coefficients 
have been computed in~\cite{Beneke:2002rj,Ciuchini:2001vx,Franco:2002fc}. 
The dimension seven contributions are known at LO~\cite{Lenz:2013aua,Gabbiani:2004tp}. 
We define the set of operators in Section~\ref{sec:lifetimes_ops} and 
present the results for their Bag parameters in Section~\ref{sec:lifetimes_bags}. 
The updated HQE results for the $B$-meson lifetime ratios are given 
in Section~\ref{sec:lifetimes_results}.

\subsection{Operators and matrix elements\label{sec:lifetimes_ops}}

\begin{figure}
 \begin{center}
    \includegraphics[width=0.4\textwidth]{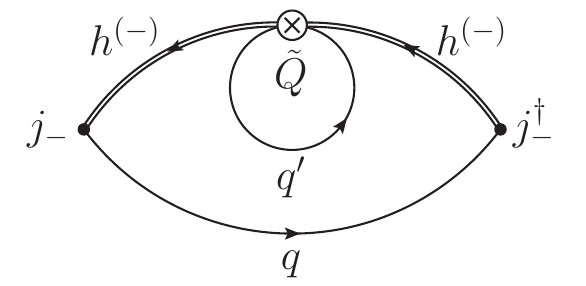}
  \caption{\label{fig:ThreePointCorrelatorEye}
  Leading order eye contraction.}
 \end{center}
\end{figure}

The following QCD operators enter at dimension six: 
\begin{eqnarray}
 Q_1^q & = & \bar{b}\gamma_\mu(1-\gamma^5)q\,\,\bar{q}\gamma^\mu(1-\gamma^5)b,\hspace{1.0cm} T_1^q = \bar{b}\gamma_\mu(1-\gamma^5)T^Aq\,\,\bar{q}\gamma^\mu(1-\gamma^5)T^Ab, \nonumber\\
 Q_2^q & = & \bar{b}(1-\gamma^5)q\,\,\bar{q}(1+\gamma^5)b,\hspace{1.85cm} T_2^q = \bar{b}(1-\gamma^5)T^Aq\,\,\bar{q}(1+\gamma^5)T^Ab. 
 \label{eq:Lifetimes_QCD_operators}
\end{eqnarray}
On the HQET side they match onto 
\begin{eqnarray}
 \tilde{Q}_1^q & = & \bar{h}\gamma_\mu(1-\gamma^5)q\,\,\bar{q}\gamma^\mu(1-\gamma^5)h,\hspace{1.0cm} \tilde{T}_1^q = \bar{h}\gamma_\mu(1-\gamma^5)T^Aq\,\,\bar{q}\gamma^\mu(1-\gamma^5)T^Ah, \nonumber\\
 \tilde{Q}_2^q & = & \bar{h}(1-\gamma^5)q\,\,\bar{q}(1+\gamma^5)h,\hspace{1.85cm} \tilde{T}_2^q = \bar{h}(1-\gamma^5)T^Aq\,\,\bar{q}(1+\gamma^5)T^Ah.
 \label{eq:Lifetimes_HQET_operators}
\end{eqnarray}
Our basis of evanescent operators and the results of the matching 
computation can be found in Appendix~\ref{sec:appendix_lifetimes}. 
We only consider the isospin-breaking combinations of operators 
\begin{equation}
 Q_i = Q_i^u - Q_i^d, \hspace{1cm} T_i = T_i^u - T_i^d, 
 \label{eq:Lifetimes_blind_ops}
\end{equation}
and their analogues in HQET. This implies that the eye contractions 
displayed in Figure~\ref{fig:ThreePointCorrelatorEye} cancel in the 
limit of exact isospin symmetry. 

The matrix elements are 
\begin{eqnarray}
  \braket{Q_i(\mu)} = A_i\, f_B^2M_B^2\, B_i(\mu), \hspace{1cm}
  \braket{T_i(\mu)} = A_i\, f_B^2M_B^2\, \epsilon_i(\mu),
  \label{eq:Lifetimes_Bags_QCD}
\end{eqnarray}
where $\braket{Q} = \braket{B^-|Q|B^-}$, the coefficients read 
\begin{equation}
  A_1 = 1, \hspace{1cm} A_{2} = \frac{M_B^2}{(m_b+m_q)^2}, 
 \label{eq:Lifetimes_AQi_QCD}
\end{equation}
and $B_{i}=1,\,\epsilon_{i}=0$ corresponds to the VSA approximation. 
Similarly we obtain for the HQET operators 
\begin{eqnarray}
 \hm{\langle}\hspace{-0.2cm}\hm{\langle}\tilde{Q}_i(\mu)\hm{\rangle}\hspace{-0.2cm}\hm{\rangle} = \tilde{A}_{i}\, F^2(\mu)\, \tilde{B}_{i}(\mu), \hspace{1cm}
 \hm{\langle}\hspace{-0.2cm}\hm{\langle}\tilde{T}_i(\mu)\hm{\rangle}\hspace{-0.2cm}\hm{\rangle} = \tilde{A}_{i}\, F^2(\mu)\, \tilde{\epsilon}_{i}(\mu), 
 \label{eq:Lifetimes_Bags_HQET}
\end{eqnarray}
where 
\begin{equation}
  \tilde{A}_1 = 1, \hspace{1cm} \tilde{A}_{2} = 1.
 \label{eq:Lifetimes_AQi_HQET}
\end{equation}

\subsection{Results for the spectral functions and bag parameters \label{sec:lifetimes_bags}}

For the $\Delta B=0$ operators we use the same conventions for the 
decomposition of the three-point correlator and the sum rule as 
for the $\Delta B=2$ operators above. 
We obtain for the double discontinuities of the non-factorizable 
contributions 
\begin{eqnarray}
 r_{\tilde{Q}_i}(x,L_\omega) & = & 0, \nonumber\\
 r_{\tilde{T}_1}(x,L_\omega) & = & -8+\frac{a_1}{8}+\frac{2 \pi ^2}{3}-\frac{3}{2}L_\omega-\frac{1}{4}\phi(x), \nonumber\\
 r_{\tilde{T}_2}(x,L_\omega) & = & -\frac{29}{4}+\frac{a_2}{8}+\frac{2 \pi ^2}{3}-\frac{3}{2}L_\omega-\frac{1}{4}\phi(x). 
 \label{eq:Lifetimes_r}
\end{eqnarray}
The leading condensate contributions have been determined in~\cite{Baek:1998vk}. 
From their results we deduce that 
\begin{eqnarray}
 \rho_{\tilde{Q}_i}^\text{cond}(\omega_1,\omega_2)  & = & 0 + \dots, \nonumber\\
 \rho_{\tilde{T}_1}^\text{cond}(\omega_1,\omega_2)  & = & \frac{\Braket{g_s \bar{q}\sigma_{\mu\nu}G^{\mu\nu}q}}{128\pi^2} \, \left[\delta(\omega_1)+\delta(\omega_2)\right] + \dots, \nonumber\\
 \rho_{\tilde{T}_2}^\text{cond}(\omega_1,\omega_2)  & = & -\frac{1}{64\pi^2} \,\left[\Braket{\frac{\alpha_s}{\pi}G^2} + \Braket{g_s \bar{q}\sigma_{\mu\nu}G^{\mu\nu}q} \left[\delta(\omega_1)+\delta(\omega_2)\right]\right] + \dots, 
 \label{eq:Lifetimes_cond}
\end{eqnarray}
where the dots indicate factorizable contributions, $\alpha_s$ 
corrections and contributions from condensates of dimension six 
and higher. To determine the condensate contributions to the HQET 
parameters we have used the traditional form of the sum rule, because 
the appearance of the $\delta$-functions obviously prevents the 
application of a weight function analogous to~\eqref{eq:weight_function}. 
We find 
\begin{eqnarray}
 \Delta \tilde{B}_i^\text{ cond}(1.5\text{ GeV})  & = & 0.000\pm0.002, \nonumber \\
 \Delta \tilde{\epsilon}_1^\text{ cond}(1.5\text{ GeV}) & = & -0.005\pm0.003, \nonumber \\
 \Delta \tilde{\epsilon}_2^\text{ cond}(1.5\text{ GeV}) & = & +0.006\pm0.004. 
\end{eqnarray}
The associated errors were determined from an uncertainty of $\pm0.002$ 
for missing higher-dimensional condensates, variations of the Borel 
parameters and the continuum cutoff and the uncertainty in the condensates  
\begin{equation}
 \Braket{\frac{\alpha_s}{\pi}G^2} = (0.012\pm0.006)\text{ GeV}^4, \hspace{1cm} \Braket{g_s \bar{q}\sigma_{\mu\nu}G^{\mu\nu}q} = (-0.011\pm0.002)\text{ GeV}^5. 
\end{equation}
We note that our results for the contributions of the condensate corrections 
to the deviation of the Bag parameters from the VSA are much smaller than those 
of~\cite{Baek:1998vk}. This is mostly due to the choice of the Borel parameter. 
We use $t\sim1\,\text{GeV}$ where the sum rule is stable against variations 
of the Borel parameter, while the Borel region of~\cite{Baek:1998vk} translates 
to $t = (0.35-0.5)\,\text{GeV}$ where the sum rule becomes unstable as can be 
seen in their plots. Our choice is also preferred by other modern sum rule 
analyses~\cite{Mannel:2007am,Mannel:2011zza,Lucha:2011zp,Gelhausen:2013wia}. 

Following analysis strategy for the perturbative contributions 
described for the $\Delta B=2$ Bag parameters in 
Section~\ref{sec:results_details}, we find the HQET Bag parameters 
\begin{equation}
\begin{aligned}
 \tilde{B}_1 (1.5\text{ GeV})        & = \hspace{0.33cm}1.000\,_{-0.020}^{+0.020} = \hspace{0.33cm}1.000\,_{-0.000}^{+0.000}(\Lbar)\,_{-0.020}^{+0.020}(\text{intr.})\,_{-0.002}^{+0.002}(\text{cond.})\,_{-0.001}^{+0.000}(\mu_\rho),\vspace*{0.1cm}\\
 \tilde{B}_2 (1.5\text{ GeV})        & = \hspace{0.33cm}1.000\,_{-0.020}^{+0.020} = \hspace{0.33cm}1.000\,_{-0.000}^{+0.000}(\Lbar)\,_{-0.020}^{+0.020}(\text{intr.})\,_{-0.002}^{+0.002}(\text{cond.})\,_{-0.001}^{+0.000}(\mu_\rho),\vspace*{0.1cm}\\
 \tilde{\epsilon}_1 (1.5\text{ GeV}) & =               -0.016\,_{-0.022}^{+0.021} =               -0.016\,_{-0.008}^{+0.007}(\Lbar)\,_{-0.020}^{+0.020}(\text{intr.})\,_{-0.003}^{+0.003}(\text{cond.})\,_{-0.003}^{+0.003}(\mu_\rho),\vspace*{0.1cm}\\
 \tilde{\epsilon}_2 (1.5\text{ GeV}) & = \hspace{0.33cm}0.004\,_{-0.022}^{+0.022} = \hspace{0.33cm}0.004\,_{-0.008}^{+0.007}(\Lbar)\,_{-0.020}^{+0.020}(\text{intr.})\,_{-0.004}^{+0.004}(\text{cond.})\,_{-0.002}^{+0.002}(\mu_\rho).
 \label{eq:Lifetimes_results_dim6_HQET}
\end{aligned}
\end{equation}
where we have set $a_1=a_2=0$. 
At the considered order there is no deviation from the VSA for the Bag 
parameters of the color singlet operators, as can be seen in~\eqref{eq:Lifetimes_r} 
and \eqref{eq:Lifetimes_cond}, because the corresponding color factors vanish. 
The deviations for the color octet operators are in the range 0\,-\,2\,\% 
for scales $\mu_\rho$ between 1 and 2 GeV. In QCD we obtain 
\begin{eqnarray}
 \overline{B}_1 (\mu=\overline{m}_b(\overline{m}_b))        & = & \hspace{0.33cm}1.028\,_{-0.056}^{+0.064} 
 = \hspace{0.33cm}1.028\,_{-0.019}^{+0.019}(\text{sum rule})\,_{-0.053}^{+0.061}(\text{matching}),\vspace*{0.1cm}\nonumber\\
 \overline{B}_2 (\mu=\overline{m}_b(\overline{m}_b))        & = & \hspace{0.33cm}0.988\,_{-0.079}^{+0.087} 
 = \hspace{0.33cm}0.988\,_{-0.020}^{+0.020}(\text{sum rule})\,_{-0.077}^{+0.085}(\text{matching}),\vspace*{0.1cm}\nonumber\\
 \overline{\epsilon}_1 (\mu=\overline{m}_b(\overline{m}_b)) & = & -0.107\,_{-0.029}^{+0.028} 
 = -0.107\,_{-0.024}^{+0.023}(\text{sum rule})\,_{-0.017}^{+0.015}(\text{matching}),\vspace*{0.1cm}\nonumber\\
 \overline{\epsilon}_2 (\mu=\overline{m}_b(\overline{m}_b)) & = & -0.033\,_{-0.021}^{+0.021} 
 = -0.033\,_{-0.018}^{+0.018}(\text{sum rule})\,_{-0.011}^{+0.011}(\text{matching}).
 \label{eq:Lifetimes_results_dim6}
\end{eqnarray}
The RG evolution and the perturbative matching cause larger deviations 
from the VSA which, however, do not exceed 11\%. 
\begin{figure}[t]
 \begin{center}
    \includegraphics[width=\textwidth]{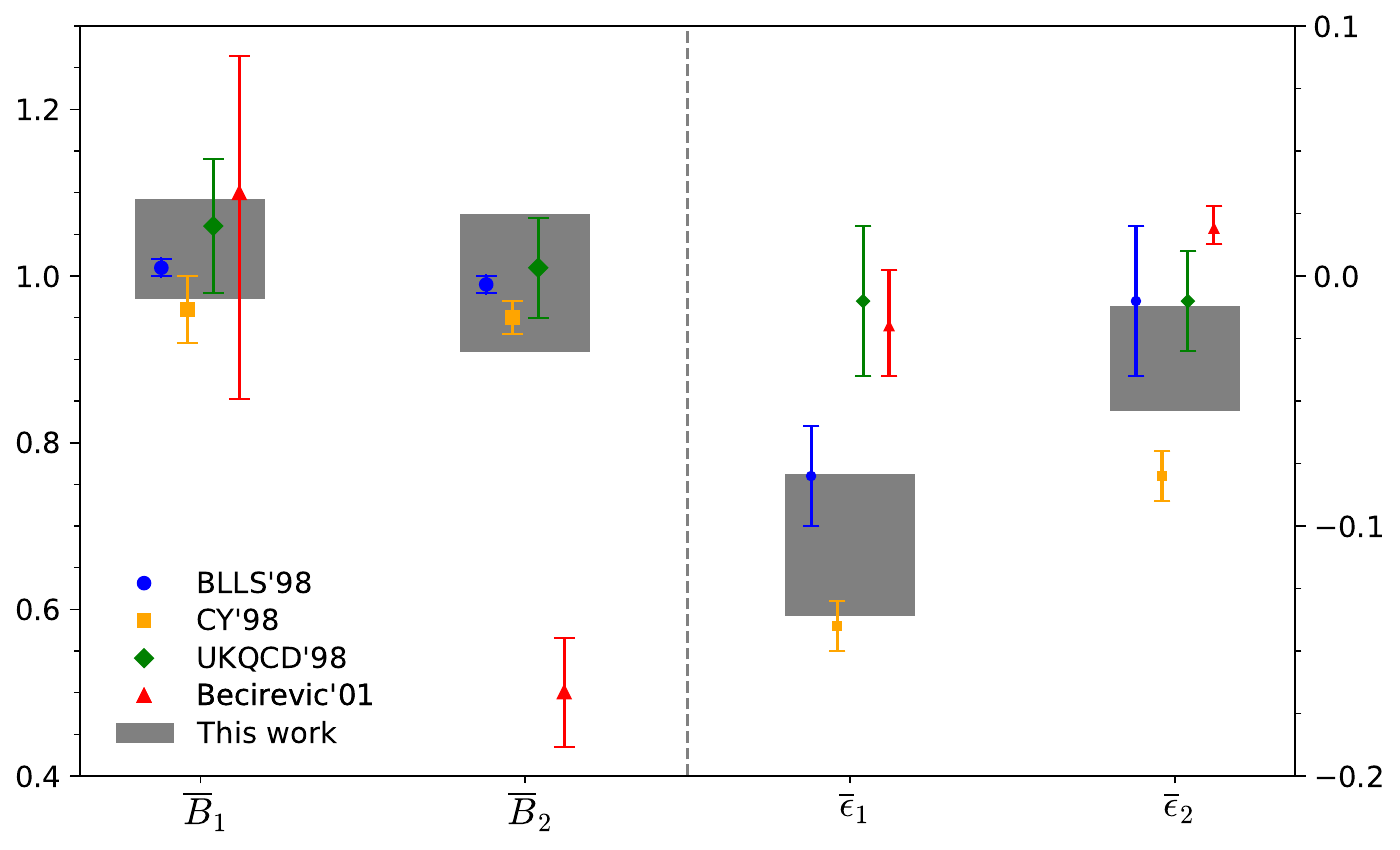}
  \caption{Comparison of our results for the $\Delta B=0$ Bag parameters at the 
  scale $\overline{m}_b(\overline{m}_b)$ to the HQET sum rule results 
  BLLS'98~\cite{Baek:1998vk} and CY'98~\cite{Cheng:1998ia},
  and the lattice values of UKQCD'98~\cite{DiPierro:1998ty} and  
  Becirevic'01~\cite{Becirevic:2001fy}.}
  \label{fig:DelB0_comparison}
 \end{center}
\end{figure}
In Figure~\ref{fig:DelB0_comparison} we compare our results to previous 
ones from sum rules~\cite{Baek:1998vk,Cheng:1998ia} and the 
lattice~\cite{DiPierro:1998ty,Becirevic:2001fy}. The results 
of~\cite{Baek:1998vk,Cheng:1998ia,DiPierro:1998ty} were obtained within HQET. 
For the comparison we match their results to QCD at tree level while 
expanding factors of $\tilde{A}_i/\overline{A}_Q(\overline{m}_b(\overline{m}_b))$ 
in $1/m_b$. As discussed in Section~\ref{sec:results_details} this 
effectively includes $1/m_b$ corrections in the VSA approximation. 

The $\overline{B}_i$ are in good agreement, with the exception of the value 
for $\overline{B}_2$ from~\cite{Becirevic:2001fy}, which differs from the other 
results and the VSA by a factor of about two. While the other sum rule results 
for the $\overline{\epsilon}_i$ agree reasonably well with ours, the lattice 
results for $\overline{\epsilon}_1$ show significantly smaller deviations 
from the VSA. The similarity between the sum rule results~\cite{Baek:1998vk,Cheng:1998ia} 
and ours appears to be mostly coincidental. As discussed above, we find that 
the bulk of the deviation from the VSA in the $\overline{\epsilon}_i$ is due 
to the RG running and matching, while the latter was not considered 
in~\cite{Baek:1998vk,Cheng:1998ia}. In their analyses, there is instead a sizeable 
deviation at the hadronic scale, originating from the condensate contributions. 
In comparison with~\cite{Baek:1998vk} we find that this is due to the choice 
of very small values of the Borel parameter which lie outside of the stability 
region as discussed above. 
The assessment of the origin of the smallness of the lattice 
results~\cite{DiPierro:1998ty,Becirevic:2001fy} for the $\overline{\epsilon}_i$ 
is beyond the scope of this work. Many of the approximations made 
in~\cite{DiPierro:1998ty,Becirevic:2001fy}, like quenching, have since been 
reappraised and a comparison with a state-of-the art lattice simulation is 
required.

\subsection{Results for the lifetime ratios \label{sec:lifetimes_results}}

Using our results~\eqref{eq:Lifetimes_results_dim6} for the dimension-six 
Bag parameters and the VSA for the dimension-seven Bag parameters defined 
in~\cite{Lenz:2013aua}, $\rho_i=1\pm1/12$, $\sigma_i=0\pm1/6$, 
\begin{equation}
 \begin{array}{lll}
 \left.\frac{\tau(B^+)}{\tau(B^0)}\right|_\text{exp}             & = 1.076\pm0.004, \hspace{-1cm}& \vspace*{0.1cm}\\
 \left.\frac{\tau(B^+)}{\tau(B^0)}\right|_{\overline{\text{MS}}} & = 1.078_{-0.023}^{+0.021} & = 1.078\,_{-0.019}^{+0.020}\,\,(\text{had.})\,_{-0.011}^{+0.002}\,(\text{scale})\pm 0.006\,(\text{param.}), \vspace{0.1cm}\\
 \left.\frac{\tau(B^+)}{\tau(B^0)}\right|_\text{PS}              & = 1.082_{-0.026}^{+0.022} & = 1.082\pm 0.021\,(\text{had.})\,_{-0.015}^{+0.000}\,(\text{scale})\pm 0.006\,(\text{param.}), \vspace{0.1cm}\\
 \left.\frac{\tau(B^+)}{\tau(B^0)}\right|_\text{1S}              & = 1.082_{-0.028}^{+0.023} & = 1.082\,_{-0.021}^{+0.022}\,\,(\text{had.})\,_{-0.017}^{+0.001}\,(\text{scale})\,_{-0.006}^{+0.007}\,\,(\text{param.}), \vspace{0.1cm}\\
 \left.\frac{\tau(B^+)}{\tau(B^0)}\right|_\text{kin}             & = 1.081_{-0.027}^{+0.022} & = 1.081\pm 0.021\,(\text{had.})\,_{-0.016}^{+0.001}\,(\text{scale})\pm 0.006\,(\text{param.}), 
 \end{array}
\end{equation}
we find excellent agreement with the experimental value and very good consistency 
between different mass schemes. The biggest contributions to the total 
uncertainty are still from the hadronic matrix elements, specifically from 
$\epsilon_1$ with $\pm0.015$ and $\sigma_3$ with $\pm0.013$. In the future, 
they can be reduced with an independent determination of the dimension-six 
Bag parameters and a sum-rule determination of the dimension-seven Bag parameters. 

We also update the prediction for the lifetime ratio $\tau(B_s^0)/\tau(B^0)$ 
in the $\overline{\text{MS}}$ scheme using Eq.~(117) from~\cite{Lenz:2015dra}: 
\begin{equation}
 \begin{array}{lll}
 \left.\frac{\tau(B_s^0)}{\tau(B^0)}\right|_\text{exp}             & = 0.994\pm 0.004, & \vspace*{0.1cm}\\
 \left.\frac{\tau(B_s^0)}{\tau(B^0)}\right|_{\overline{\text{MS}}} & = 1.0007\pm 0.0025 & \vspace*{0.1cm}\\
 & = 1.0007\pm 0.0014\,(\text{had.})\,\pm 0.0006\,(\text{scale})\,\pm 0.0020\,(1/m_b^4), &
 \end{array}
\end{equation}
where we have added an uncertainty estimate for the spectator effects at order 
$1/m_b^4$ which have not been considered in~\cite{Lenz:2015dra}. With respect 
to last year~\cite{Jubb:2016mvq}, the difference between the theory prediction 
and the experimental value for $\tau(B_s^0)/\tau(B^0)$ is reduced from 
$2.5\,\sigma$ to $1.4\,\sigma$.


\section{\boldmath Matrix elements for charm and the $D^+-D^0$ lifetime ratio\label{sec:charm}}

The HQET sum rule analysis can easily be adapted to the charm sector. 
It is common to quote the matrix elements for the charm sector at the scale 
3 GeV instead of the charm-quark mass, 
see~\cite{Carrasco:2014uya,Carrasco:2015pra,Bazavov:2017weg}, and we adopt 
that convention for ease of comparison. Consequently we also use 3 GeV 
as the central matching scale. In the error analysis it is varied between 
2 and 4 GeV. To account for the lower value of charm-quark mass we assume 
that the uncertainty due to power corrections is 0.03 instead of 0.01 
for the bottom sector. Otherwise we use the same analysis strategy as in 
the bottom sector which is outlined in Section~\ref{sec:results_details}. 

\subsection{\boldmath Matrix elements for $D$ mixing\label{sec:Dmixing}}

The latest lattice QCD study~\cite{Bazavov:2017weg} for $D$ mixing only 
gives results for the matrix elements and not for the Bag parameters. 
We do the same here and obtain, using the value of the $D$-meson decay 
constant from Appendix~\ref{sec:Uncertainties}, 
\begin{eqnarray}
  \Braket{Q_1 (3\,\text{GeV})}/\text{GeV}^4 & = & 0.265\,_{-0.021}^{+0.024} = 0.265\,_{-0.010}^{+0.006}\,(\text{s.r.})\,_{-0.014}^{+0.019}\,(\text{matching})\,_{-0.012}^{+0.013}\,(f_D),\nonumber\\
 -\Braket{Q_2 (3\,\text{GeV})}/\text{GeV}^4 & = & 0.502\,_{-0.092}^{+0.124} = 0.502\,_{-0.078}^{+0.094}\,(\text{s.r.})\,_{-0.044}^{+0.076}\,(\text{matching})\,_{-0.023}^{+0.024}\,(f_D),\nonumber\\
  \Braket{Q_3 (3\,\text{GeV})}/\text{GeV}^4 & = & 0.135\,_{-0.029}^{+0.037} = 0.135\,_{-0.026}^{+0.031}\,(\text{s.r.})\,_{-0.010}^{+0.019}\,(\text{matching})\,_{-0.006}^{+0.006}\,(f_D),\nonumber\\
  \Braket{Q_4 (3\,\text{GeV})}/\text{GeV}^4 & = & 0.792\,_{-0.122}^{+0.175} = 0.792\,_{-0.093}^{+0.116}\,(\text{s.r.})\,_{-0.070}^{+0.125}\,(\text{matching})\,_{-0.037}^{+0.038}\,(f_D),\nonumber\\
  \Braket{Q_5 (3\,\text{GeV})}/\text{GeV}^4 & = & 0.340\,_{-0.039}^{+0.060} = 0.340\,_{-0.021}^{+0.027}\,(\text{s.r.})\,_{-0.029}^{+0.051}\,(\text{matching})\,_{-0.016}^{+0.016}\,(f_D).\nonumber\\
 \label{eq:DelC2_results}
\end{eqnarray}
The relative uncertainties in the charm sector are consistently larger 
than those in the bottom sector because of larger perturbative corrections 
due to a larger value of $\alpha_s$ at the smaller scales and larger 
power corrections. This effect is most pronounced for $Q_2$, $Q_4$ 
and $Q_5$ where the relative uncertainty is larger by a factor of 
order two. In the matrix elements we have an additional uncertainty 
from the value of the decay constant which is added in quadrature. 

\begin{figure}[t]
 \begin{center}
    \includegraphics[width=\textwidth]{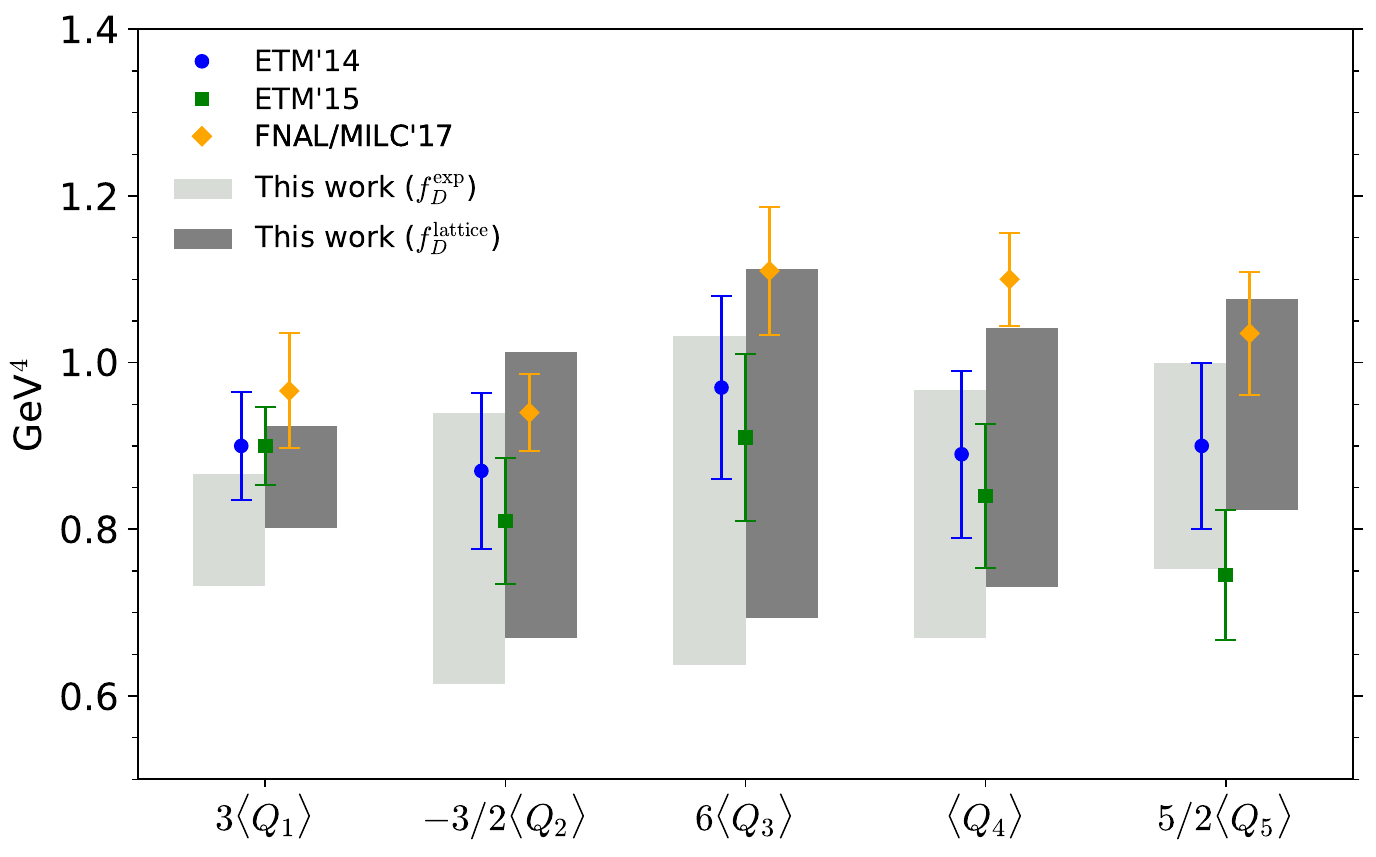}
  \caption{Comparison of our results for the $\Delta C=2$ matrix elements 
  at the scale 3 GeV to the lattice values of ETM'14~\cite{Carrasco:2014uya}, 
  ETM'15~\cite{Carrasco:2015pra} and FNAL/MILC'17~\cite{Bazavov:2017weg}. 
  The values for the matrix elements of the ETM collaboration are extracted 
  from Figure~16 of~\cite{Bazavov:2017weg}.}
  \label{fig:DelC2_comparison}
 \end{center}
\end{figure}

We compare our results to those from the lattice in Figure~\ref{fig:DelC2_comparison}. 
There is a consistent hierarchy with decreasing values from the results 
of the FNAL/MILC collaboration~\cite{Bazavov:2017weg}, those of the ETM 
collaboration~\cite{Carrasco:2014uya,Carrasco:2015pra} and ours. The only 
exception is the value of $\Braket{Q_5}$ from~\cite{Carrasco:2015pra} which 
lies below ours. If we use the lattice average~\cite{Olive:2016xmw} for the 
decay constant $f_D^\text{lattice} = (211.9\pm1.1)\,$MeV in place of the 
experimental average $f_D^\text{exp} = (203.7\pm4.8)\,$MeV~\cite{Olive:2016xmw}, 
we find very good agreement between our results and those of ETM and the 
remaining differences with respect to the FNAL/MILC results are comfortably 
below two sigmas. We prefer the experimental average of the decay constant 
since it is in significantly better agreement with recent sum rule 
results~\cite{Wang:2015mxa,Gelhausen:2013wia,Narison:2012xy,Lucha:2011zp}. 
On the other hand, using the lattice value yields a more meaningful comparison 
with the lattice results since the quantities we determine with the sum rule 
are the Bag parameters and the decay constant cancels out in the comparison 
if the same value is used on both sides. We therefore conclude that our 
sum rule results for the non-factorizable contributions to the Bag parameters 
are in good agreement with lattice simulations. An investigation of the 
differences in the numerical values of the decay constant is beyond the scope 
of this work. 

\subsection{\boldmath Matrix elements for $D$ lifetimes and $\tau(D^+)/\tau(D^0)$\label{sec:Dlifetimes}}

Our results for the $\Delta C=0$ Bag parameters are 
\begin{eqnarray}
 \overline{B}_1 (3\,\text{GeV}) & = & \hspace{0.33cm}0.902\,_{-0.051}^{+0.077} 
 = \hspace{0.33cm}0.902\,_{-0.018}^{+0.018}\,(\text{sum rule})\,_{-0.048}^{+0.075}\,(\text{matching}),\nonumber\\
 \overline{B}_2 (3\,\text{GeV}) & = & \hspace{0.33cm}0.739\,_{-0.073}^{+0.124} 
 = \hspace{0.33cm}0.739\,_{-0.015}^{+0.015}\,(\text{sum rule})\,_{-0.072}^{+0.123}\,(\text{matching}),\nonumber\\
 \overline{\epsilon}_1 (3\,\text{GeV}) & = & -0.132\,_{-0.046}^{+0.041} 
 = -0.132\,_{-0.026}^{+0.025}\,(\text{sum rule})\,_{-0.038}^{+0.033}\,(\text{matching}),\nonumber\\
 \overline{\epsilon}_2 (3\,\text{GeV}) & = & -0.005\,_{-0.032}^{+0.032} 
 = -0.005\,_{-0.012}^{+0.011}\,(\text{sum rule})\,_{-0.030}^{+0.030}\,(\text{matching}).
 \label{eq:Lifetimes_results_dim6_charm}
\end{eqnarray}
While the uncertainties in $B_{1,2}$ are similar to those in the $B$ sector 
we find that those in $\epsilon_{1,2}$ are larger by about 50\%. The latter 
ones are dominated by the non-factorizable power correction and the intrinsic 
sum rule errors which are both based on somewhat ad-hoc estimates. Thus, our 
values for the uncertainties of $\epsilon_{1,2}$ should be taken with a 
grain of salt and lattice results for the $\Delta C=0$ Bag parameters could 
provide an important consistency check. Alternatively, one could also improve 
the dominant error due to non-factorizable $1/m_c$ corrections by performing 
the operator matching up to the order $1/m_c$ and determine the matrix elements 
of the subleading HQET operators using sum rules. 

We update our result for the $D$-meson lifetime ratio from~\cite{Lenz:2013aua} 
using the dimension six Bag parameters~\eqref{eq:Lifetimes_results_dim6_charm} 
and the VSA $\rho_i=1\pm1/12$, $\sigma_i=0\pm1/6$ for the dimension-seven Bag 
parameters. We have converted the $\overline{\text{MS}}$ value of the charm-quark 
mass to the PS mass at $\mu_f=1\,$GeV and the 1S mass at four-loop accuracy 
using \texttt{RunDec}. The kinetic mass at the scale 1\,GeV is determined with 
two-loop accuracy using an unpublished version of the \texttt{QQbar\_Threshold} 
code~\cite{Beneke:2016kkb,Beneke:2017rdn}. The central value for the scales 
$\mu_1$ and $\mu_0$ is fixed to $1.5\,$GeV for all mass schemes and varied 
between 1 and 3 GeV. We find 
\begin{align}
 & \left.\frac{\tau(D^+)}{\tau(D^0)}\right|_\text{exp}             \hspace{-0.7cm}& = 2.536\pm0.019, \hspace{-1cm}& \vspace*{0.1cm} \nonumber\\
 & \left.\frac{\tau(D^+)}{\tau(D^0)}\right|_{\overline{\text{MS}}} \hspace{-0.7cm}& = 2.61_{-0.77}^{+0.72} & = 2.61\,_{-0.66}^{+0.70}\,(\text{had.})\,_{-0.38}^{+0.12}\,(\text{scale})\pm 0.09\,(\text{param.}), \vspace{0.1cm} \nonumber\\
 & \left.\frac{\tau(D^+)}{\tau(D^0)}\right|_{\text{PS}}            \hspace{-0.7cm}& = 2.70_{-0.82}^{+0.74} & = 2.70\,_{-0.68}^{+0.72}\,(\text{had.})\,_{-0.45}^{+0.11}\,(\text{scale})\pm 0.10\,(\text{param.}), \vspace{0.1cm}\\
 & \left.\frac{\tau(D^+)}{\tau(D^0)}\right|_{\text{1S}}            \hspace{-0.7cm}& = 2.56_{-0.99}^{+0.81} & = 2.56\,_{-0.74}^{+0.78}\,(\text{had.})\,_{-0.65}^{+0.22}\,(\text{scale})\pm 0.10\,(\text{param.}), \vspace{0.1cm} \nonumber\\
 & \left.\frac{\tau(D^+)}{\tau(D^0)}\right|_{\text{kin}}           \hspace{-0.7cm}& = 2.53_{-0.76}^{+0.72} & = 2.53\,_{-0.66}^{+0.70}\,(\text{had.})\,_{-0.37}^{+0.13}\,(\text{scale})\pm 0.10\,(\text{param.}), \nonumber
 \end{align}
which is in very good agreement. The various mass schemes are all consistent 
and we again take the PS result as our preferred value. The dominant sources 
of uncertainties are the Bag parameters $\epsilon_1$ and $\sigma_3$ which both 
contribute $\pm0.5$ to the error budget of the lifetime ratio. Both errors can 
be reduced in the future with a lattice determination of the dimensions-six 
matrix elements and a sum-rule determination of the dimension-seven Bag parameters, 
respectively. In the PS scheme, the radiative and power corrections are of 
the order $+27\%$ and $-34\%$, respectively, which indicates good convergence 
behaviour. We therefore conclude that the HQE provides a good description 
of the lifetime ratio $\tau(D^+)/\tau(D^0)$.


\section{Conclusions\label{sec:conclusion}}

We have determined the matrix elements of the dimension six $\Delta F = 0,2$ operators 
for the bottom and charm sector using HQET sum rules. Our findings for the $\Delta F = 2$ 
matrix elements are in good agreement with recent lattice~\cite{Dalgic:2006gp,Carrasco:2013zta,Bazavov:2016nty,
Carrasco:2014uya,Carrasco:2015pra,Bazavov:2017weg} and sum rule~\cite{Grozin:2016uqy} results. 
\begin{figure}[t]
 \begin{center}
    \includegraphics[width=0.4855\textwidth]{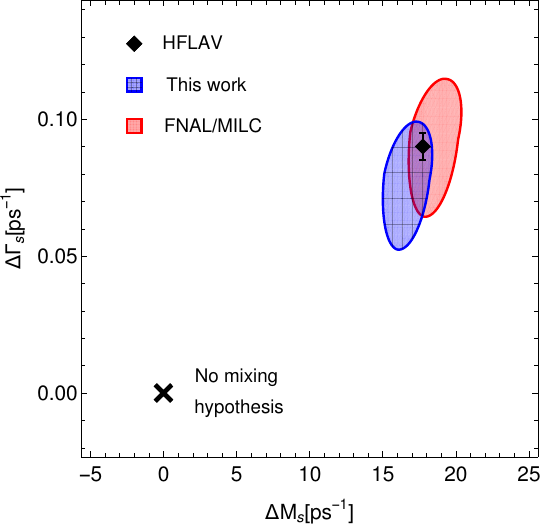}
    \hfill
    \includegraphics[width=0.4745\textwidth]{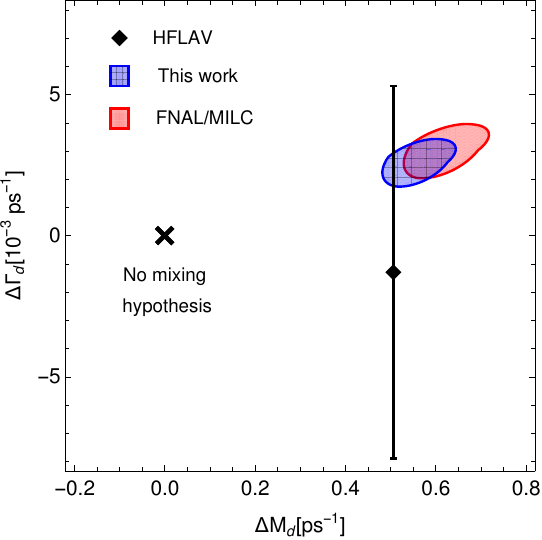}
  \caption{Comparison of our predictions for the mass and decay rate difference 
  in the $B_s$ (left) and $B_d$ (right) system with the present experimental 
  averages (error bars). We also show the results obtained with the lattice results 
  of~\cite{Bazavov:2016nty} for $f_{B_q}^2\overline{B}_{Q_i}$ and the matrix element 
  $\langle R_0\rangle$ and the values given in Appendix~\ref{sec:Uncertainties} for the other 
  input parameters. The PS mass scheme for the bottom quarks has been used in both cases.}
  \label{fig:MixingSMvsExp}
 \end{center}
\end{figure}
Our $\Delta F = 0$ results are the first state-of-the-art values for the matrix elements 
required for $B$ and $D$ meson lifetime ratios. 
\begin{figure}[t]
 \begin{center}
    \includegraphics[width=\textwidth]{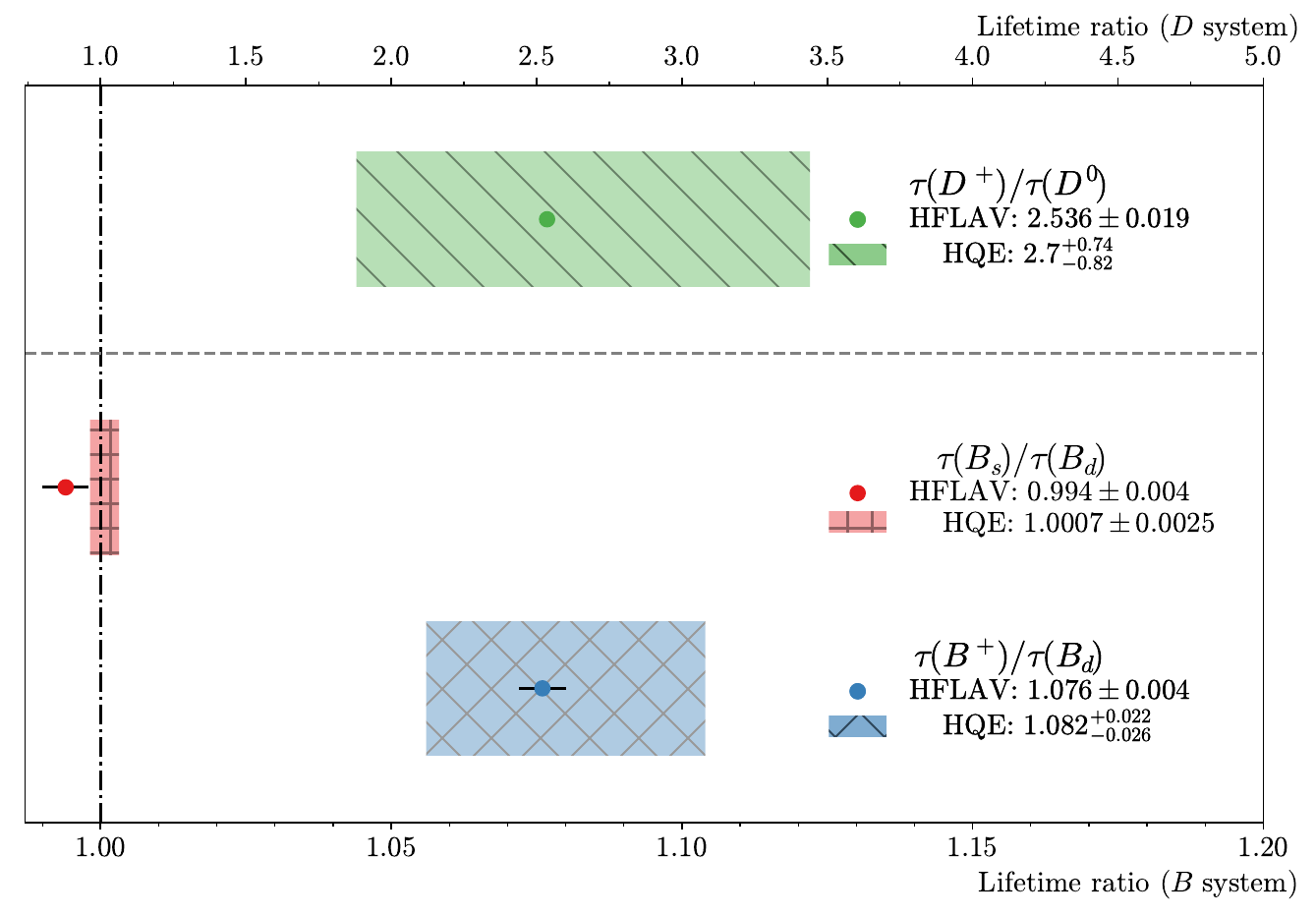}
  \caption{Comparison of our predictions for the lifetime ratios of heavy mesons
  with the present experimental averages. }
  \label{fig:LifetimesSMvsExp}
 \end{center}
\end{figure}
The uncertainties in our analyses for the Bag parameters are similar to those of 
recent lattice determinations in the $B$ sector and somewhat larger in the $D$ sector. 
This suggests that the uncertainty of the $\Delta C=0$ matrix elements could be 
reduced by a lattice simulation. In most cases, the dominant errors in our approach 
stem from the matching of QCD to HQET operators, see Appendix~\ref{sec:Uncertainties}. 
These could be reduced substantially by performing the matching calculation at NNLO. 
Some first steps towards this goal have recently been taken in~\cite{Grozin:2017uto}. 
Consequently, in the future, sum rules will continue to be competitive with lattice 
simulations in the determination of four-quark operators. 

Our predictions for the mixing observables and lifetime ratios in the $B$ sector are 
in good agreement with the experimental averages as summarized in Figures~\ref{fig:MixingSMvsExp} 
and~\ref{fig:LifetimesSMvsExp}. In particular, the small tensions~\cite{Bazavov:2016nty,Jubb:2016mvq} 
that follow from using the FNAL/MILC results~\cite{Bazavov:2016nty} for the matrix 
elements are not confirmed by our results. We note that the predictions based on 
matrix elements from sum rules and from lattice simulations are compatible and lead 
to overall uncertainties of the same size. Taking the naive average of the Bag 
parameters, the relative uncertainties of the mass and decay rate difference are, 
however, only reduced by about $9\%$ and $6\%$, respectively, because other sources 
of uncertainties, like e.g. the matrix elements of dimension-seven operators, are dominant. 

We find that the experimental value for the lifetime ratio $\tau(D^+)/\tau(D^0)$ 
can be reproduced within the HQE. This is a strong indication that the HQE does 
not break down in the charm sector. However, due to sizeable hadronic uncertainties, 
we cannot exclude large duality violations at the level of 20-30\% yet. On the other 
hand, the $D$-mixing observables are very sensitive to duality violations and might 
offer a handle on a better quantitative understanding of these effects~\cite{Bigi:2000wn}. 

Our comprehensive study demonstrates that the HQET sum rules for hadronic four-quark 
matrix elements provide a competitive alternative to lattice simulations. Due to 
completely different systematics they facilitate powerful independent checks of 
lattice results. Sum rules can also be applied to obtain the matrix elements 
of the subleading dimension-seven operators, which have never been determined using 
lattice simulations. This is crucial to achieve a substantial reduction of the 
current theoretical uncertainties.

\subsection*{Acknowledgements}

We are grateful to V. Braun for illuminating discussions and to A.~Grozin and 
A.~Pivovarov for pointing out to us that some of the color factors in Eqs.~\eqref{eq:r_def},
\eqref{eq:weight_function} and \eqref{eq:B_NoInt} were set to their QCD values in an earlier 
version of the manuscript. We wish to thank G.~Tetlalmatzi-Xolocotzi for poiting out that 
a wrong input value was used in our numerical code for $\Delta M_q$ and $a_\text{sl}^q$. 
This work was supported by the STFC through the IPPP grant.

\appendix


\section{Basis of evanescent operators and ADMs\label{sec:EvOps_and_ADMs}}

\subsection{\boldmath $\Delta B=2$ operators\label{sec:appendix_mixing}}

Our choice of basis for the evanescent operators is given by 
\begin{eqnarray}
 E_1 & = & \bar{b}_i\gamma_\mu(1-\gamma^5)q_j\,\,\bar{b}_j\gamma^\mu(1-\gamma^5)q_i - Q_1, \nonumber\\
 E_2 & = & \bar{b}_i\gamma_\mu\gamma_\nu(1-\gamma^5)q_i\,\,\bar{b}_j\gamma^\mu\gamma^\nu(1-\gamma^5)q_j - (8-4\epsilon)Q_2 - (8-8\epsilon)Q_3, \nonumber\\ 
 E_3 & = & \bar{b}_i\gamma_\mu\gamma_\nu(1-\gamma^5)q_j\,\,\bar{b}_j\gamma^\mu\gamma^\nu(1-\gamma^5)q_i - (8-8\epsilon)Q_2 - (8-4\epsilon)Q_3, \nonumber\\
 E_4 & = & \bar{b}_i\gamma_\mu\gamma_\nu\gamma_\rho(1-\gamma^5)q_i\,\,\bar{b}_j\gamma^\mu\gamma^\nu\gamma^\rho(1-\gamma^5)q_j - (16-4\epsilon)Q_1, \nonumber\\
 E_5 & = & \bar{b}_i\gamma_\mu\gamma_\nu\gamma_\rho(1-\gamma^5)q_j\,\,\bar{b}_j\gamma^\mu\gamma^\nu\gamma^\rho(1-\gamma^5)q_i - (16-4\epsilon)(Q_1+E_1), \nonumber\\
 E_6 & = & \bar{b}_i\gamma_\mu(1-\gamma^5)q_i\,\,\bar{b}_j\gamma^\mu(1+\gamma^5)q_j + 2Q_5, \nonumber\\
 E_7 & = & \bar{b}_i\gamma_\mu(1-\gamma^5)q_j\,\,\bar{b}_j\gamma^\mu(1+\gamma^5)q_i + 2Q_4, \nonumber\\
 E_8 & = & \bar{b}_i\gamma_\mu\gamma_\nu(1-\gamma^5)q_i\,\,\bar{b}_j\gamma^\mu\gamma^\nu(1+\gamma^5)q_j - 4Q_4, \nonumber\\
 E_9 & = & \bar{b}_i\gamma_\mu\gamma_\nu(1-\gamma^5)q_j\,\,\bar{b}_j\gamma^\mu\gamma^\nu(1+\gamma^5)q_i - 4Q_5, 
 \label{eq:QCD_ev_operators}
\end{eqnarray}
for QCD and 
\begin{eqnarray}
 \tilde{E}_1 & = & \bar{h}_i^{\{(+)}\gamma_\mu(1-\gamma^5)q_j\,\,\bar{h}_j^{(-)\}}\gamma^\mu(1-\gamma^5)q_i - \tilde{Q}_1,\nonumber\\
 \tilde{E}_2 & = & \frac12 \tilde{Q}_1 + \tilde{Q}_2 + \bar{h}_i^{\{(+)}(1-\gamma^5)q_j\,\,\bar{h}_j^{(-)\}}(1-\gamma^5)q_i,\nonumber\\
 \tilde{E}_3 & = & \bar{h}_i^{\{(+)}\gamma_\mu\gamma_\nu(1-\gamma^5)q_i\,\,\bar{h}_j^{(-)\}}\gamma^\mu\gamma^\nu(1-\gamma^5)q_j + (4+a_1\epsilon)\tilde{Q}_1,\nonumber\\
 \tilde{E}_4 & = & \bar{h}_i^{\{(+)}\gamma_\mu\gamma_\nu(1-\gamma^5)q_j\,\,\bar{h}_j^{(-)\}}\gamma^\mu\gamma^\nu(1-\gamma^5)q_i + (4+a_1\epsilon)(\tilde{Q}_1+\tilde{E}_1),\nonumber\\
 \tilde{E}_5 & = & \bar{h}_i^{\{(+)}\gamma_\mu\gamma_\nu\gamma_\rho(1-\gamma^5)q_i\,\,\bar{h}_j^{(-)\}}\gamma^\mu\gamma^\nu\gamma^\rho(1-\gamma^5)q_j - (16+a_2\epsilon)\tilde{Q}_1,\nonumber\\
 \tilde{E}_6 & = & \bar{h}_i^{\{(+)}\gamma_\mu\gamma_\nu\gamma_\rho(1-\gamma^5)q_j\,\,\bar{h}_j^{(-)\}}\gamma^\mu\gamma^\nu\gamma^\rho(1-\gamma^5)q_i - (16+a_2\epsilon)(\tilde{Q}_1+\tilde{E}_1),\nonumber\\
 \tilde{E}_7 & = & \bar{h}_i^{\{(+)}\gamma_\mu(1-\gamma^5)q_i\,\,\bar{h}_j^{(-)\}}\gamma^\mu(1+\gamma^5)q_j + 2\tilde{Q}_5,\nonumber\\
 \tilde{E}_8 & = & \bar{h}_i^{\{(+)}\gamma_\mu(1-\gamma^5)q_j\,\,\bar{h}_j^{(-)\}}\gamma^\mu(1+\gamma^5)q_i + 2\tilde{Q}_4,\nonumber\\
 \tilde{E}_9 & = & \bar{h}_i^{\{(+)}\gamma_\mu\gamma_\nu(1-\gamma^5)q_i\,\,\bar{h}_j^{(-)\}}\gamma^\mu\gamma^\nu(1+\gamma^5)q_j  - (4+a_3\epsilon)\tilde{Q}_4,\nonumber\\
 \tilde{E}_{10} & = & \bar{h}_i^{\{(+)}\gamma_\mu\gamma_\nu(1-\gamma^5)q_j\,\,\bar{h}_j^{(-)\}}\gamma^\mu\gamma^\nu(1+\gamma^5)q_i  - (4+a_3\epsilon)\tilde{Q}_5,
 \label{eq:HQET_ev_operators}
\end{eqnarray}
for HQET. It is straightforward to verify that the evanescent operators 
vanish in four dimensions by using the Fierz identities 
\begin{equation}
\begin{aligned}
 &\left[\gamma_\mu(1\pm\gamma_5)\right]_{ij}\left[\gamma^\mu(1\pm\gamma_5)\right]_{kl}=-\left[\gamma_\mu(1\pm\gamma_5)\right]_{il}\left[\gamma^\mu(1\pm\gamma_5)\right]_{kj},\\
 &\left[1\pm\gamma_5\right]_{ij}\left[1\pm\gamma_5\right]_{kl}=\frac12\left[1\pm\gamma_5\right]_{il}\left[1\pm\gamma_5\right]_{kj}
 +\frac18\left[\sigma_{\mu\nu}(1\pm\gamma_5)\right]_{il}\left[\sigma^{\mu\nu}(1\pm\gamma_5)\right]_{kj},\\
 &\left[\gamma_\mu(1\pm\gamma_5)\right]_{ij}\left[\gamma^\mu(1\mp\gamma_5)\right]_{kl}=2\left[1\mp\gamma_5\right]_{il}\left[1\pm\gamma_5\right]_{kj},
 \label{eq:Fierz4D}
\end{aligned}
\end{equation}
and the relation 
\begin{equation}
 \gamma_\mu\gamma_\nu\gamma_\rho = g_{\mu\nu}\gamma_\rho + g_{\nu\rho}\gamma_{\mu} - g_{\mu\rho}\gamma_\nu - i\epsilon_{\mu\nu\rho\lambda}\gamma^\lambda\gamma^5. 
 \label{eq:three_gammas}
\end{equation}
A useful strategy to simplify expressions with two Dirac matrices 
is to use projection identities, e.g. 
\begin{equation}
 \bar{h}^{(\pm)}\gamma_\mu\gamma_\nu(1-\gamma^5)q = \pm\bar{h}^{(\pm)}\slashed{v}\gamma_\mu\gamma_\nu(1-\gamma^5)q,
\end{equation}
and then reduce the number of Dirac matrices with Eq.~\eqref{eq:three_gammas}. 

In the decomposition~\eqref{eq:ADM_QCD} the LO QCD ADM is  
\begin{equation}
 \gamma_{QQ}^{(0)} = \left(
\begin{array}{ccccc}
 \frac{6(N_c-1)}{N_c} & 0 & 0 & 0 & 0 \\ \vspace*{0.1cm}
 0 & -\frac{2 \left(3 N_c^2-4 N_c-1\right)}{N_c} & \frac{4N_c-8}{N_c} & 0 & 0 \\ \vspace*{0.1cm}
 0 & \frac{4 (N_c-2) (N_c+1)}{N_c} & \frac{2 (N_c+1)^2}{N_c} & 0 & 0 \\ \vspace*{0.1cm}
 0 & 0 & 0 & -\frac{6(N_c^2-1)}{N_c} & 0 \\ \vspace*{0.1cm}
 0 & 0 & 0 & -6 & \frac{6}{N_c} \\
\end{array}
\right),
\label{eq:ADM_QQ}
\end{equation}
\begin{equation}
 \gamma_{QE}^{(0)} = \left(
\begin{array}{ccccccccc}
 6 & 0 & 0 & -\frac{1}{N_c} & 1 & 0 & 0 & 0 & 0 \\ \vspace*{0.1cm}
 0 & -\frac{1}{N_c} & 1 & 0 & 0 & 0 & 0 & 0 & 0 \\ \vspace*{0.1cm}
 0 & \frac{1}{2} & \frac{N_c}{2}-\frac{1}{N_c} & 0 & 0 & 0 & 0 & 0 & 0 \\ \vspace*{0.1cm}
 0 & 0 & 0 & 0 & 0 & 0 & 0 & -\frac{1}{N_c} & 1 \\ \vspace*{0.1cm}
 0 & 0 & 0 & 0 & 0 & 0 & 0 & \frac{1}{2} & \frac{N_c}{2}-\frac{1}{N_c} \\
\end{array}
\right).
\label{eq:ADM_QE}
\end{equation}
In HQET we find 
\begin{equation}
 \tilde{\gamma}_{\tilde{Q}\tilde{Q}}^{(0)} = \left(
\begin{array}{ccccc}
 \frac{3}{N_c}-3 N_c & 0 & 0 & 0 \\ \vspace*{0.1cm}
 1+\frac{1}{N_c} & -3 N_c+4+\frac{7}{N_c} & 0 & 0 \\ \vspace*{0.1cm}
 0 & 0 & \frac{6}{N_c}-3 N_c & -3 \\ \vspace*{0.1cm}
 0 & 0 & -3 & \frac{6}{N_c}-3 N_c \\
\end{array}
\right),
\label{eq:ADM_QtilQtil}
\end{equation}
\begin{equation}
 \tilde{\gamma}_{\tilde{Q}\tilde{E}}^{(0)} = \left(
\begin{array}{cccccccccc}
 0 & 0 & 0 & 0 & -\frac{1}{4 N_c} & \frac{1}{4} & 0 & 0 & 0 & 0 \\ \vspace*{0.1cm}
 -1 & -4 & -\frac{1}{4 N_c} & \frac{1}{4} & 0 & 0 & 0 & 0 & 0 & 0 \\ \vspace*{0.1cm}
 0 & 0 & 0 & 0 & 0 & 0 & 0 & 0 & -\frac{1}{4 N_c} & \frac{1}{4} \\ \vspace*{0.1cm}
 0 & 0 & 0 & 0 & 0 & 0 & 0 & 0 & \frac{1}{4} & -\frac{1}{4 N_c} \\
\end{array}
\right).
\label{eq:ADM_QtilEtil}
\end{equation}

Our result \eqref{eq:ADM_QQ} with $N_c=3$  differs from the 
results of~\cite{Gamiz:2008sk,Monahan:2014xra} because we have 
only used the replacements implied by the basis of evanescent 
operators~\eqref{eq:QCD_ev_operators} to simplify products of 
Dirac matrices. We can reproduce their result by applying 
4-dimensional Fierz identities that relate $Q_1$, $Q_2$ and $Q_3$. 
The upper left $2\times2$ submatrix of~\eqref{eq:ADM_QtilQtil} 
agrees with~\cite{Buchalla:1996ys}. 

\subsection{\boldmath $\Delta B=0$ operators\label{sec:appendix_lifetimes}}

We define the basis of evanescent operators in QCD 
following~\cite{Beneke:2002rj}:
\begin{eqnarray}
 E_1^q & = & \bar{b}\gamma_\mu\gamma_\nu\gamma_\rho(1-\gamma^5)q\,\,\bar{q}\gamma^\rho\gamma^\nu\gamma^\mu(1-\gamma^5)b - (4-8\epsilon)Q_1^q, \nonumber\\
 E_2^q & = & \bar{b}\gamma_\mu\gamma_\nu(1-\gamma^5)q\,\,\bar{q}\gamma^\nu\gamma^\mu(1+\gamma^5)b - (4-8\epsilon)Q_2^q, \nonumber\\
 E_3^q & = & \bar{b}\gamma_\mu\gamma_\nu\gamma_\rho(1-\gamma^5)T^Aq\,\,\bar{q}\gamma^\rho\gamma^\nu\gamma^\mu(1-\gamma^5)T^Ab - (4-8\epsilon)T_1^q, \nonumber\\
 E_4^q & = & \bar{b}\gamma_\mu\gamma_\nu(1-\gamma^5)T^Aq\,\,\bar{q}\gamma^\nu\gamma^\mu(1+\gamma^5)T^Ab - (4-8\epsilon)T_2^q. 
 \label{eq:Lifetimes_QCD_ev_operators}
\end{eqnarray}
In HQET we again introduce parameters $a_{1,2}$ to keep track of 
the scheme dependence 
\begin{eqnarray}
 \tilde{E}_1^q & = & \bar{h}\gamma_\mu\gamma_\nu\gamma_\rho(1-\gamma^5)q\,\,\bar{q}\gamma^\rho\gamma^\nu\gamma^\mu(1-\gamma^5)h - (4+a_1\epsilon)\tilde{Q}_1^q, \nonumber\\
 \tilde{E}_2^q & = & \bar{h}\gamma_\mu\gamma_\nu(1-\gamma^5)q\,\,\bar{q}\gamma^\nu\gamma^\mu(1+\gamma^5)h - (4+a_2\epsilon)\tilde{Q}_2^q, \nonumber\\
 \tilde{E}_3^q & = & \bar{h}\gamma_\mu\gamma_\nu\gamma_\rho(1-\gamma^5)T^Aq\,\,\bar{q}\gamma^\rho\gamma^\nu\gamma^\mu(1-\gamma^5)T^Ah - (4+a_1\epsilon)\tilde{T}_1^q, \nonumber\\
 \tilde{E}_4^q & = & \bar{h}\gamma_\mu\gamma_\nu(1-\gamma^5)T^Aq\,\,\bar{q}\gamma^\nu\gamma^\mu(1+\gamma^5)T^Ah - (4+a_2\epsilon)\tilde{T}_2^q. 
 \label{eq:Lifetimes_HQET_ev_operators}
\end{eqnarray}
The isospin breaking combinations of the evanescent operators 
are defined in analogy to~\eqref{eq:Lifetimes_blind_ops}. 
The LO ADM in QCD takes the form  
\begin{equation}
 \gamma_{QQ}^{(0)} = \left(
\begin{array}{ccccc}
 0 & 0 & 12 & 0\\ \vspace*{0.1cm}
 0 & \frac{6}{N_c}-6 N_c & 0 & 0\\ \vspace*{0.1cm}
 3-\frac{3}{N_c^2} & 0 & -\frac{12}{N_c} & 0\\ \vspace*{0.1cm}
 0 & 0 & 0 & \frac{6}{N_c}\\
\end{array}
\right),
\label{eq:lifetimes_ADM_QQ}
\end{equation}
\begin{equation}
 \gamma_{QE}^{(0)} = \left(
\begin{array}{ccccc}
 0 & 0 & -2 & 0\\ \vspace*{0.1cm}
 0 & 0 & 0 & -2\\ \vspace*{0.1cm}
 \frac{1}{2 N_c^2}-\frac{1}{2} & 0 & \frac{2}{N_c}-\frac{N_c}{2} & 0\\ \vspace*{0.1cm}
 0 & \frac{1}{2 N_c^2}-\frac{1}{2} & 0 & \frac{2}{N_c}-\frac{N_c}{2}\\
\end{array}
\right).
\label{eq:lifetimes_ADM_QE}
\end{equation}
The HQET result is given by 
\begin{equation}
 \tilde{\gamma}_{\tilde{Q}\tilde{Q}}^{(0)} = \left(
\begin{array}{ccccc}
 \frac{3}{N_c}-3 N_c & 0 & 6 & 0\\ \vspace*{0.1cm}
 0 & \frac{3}{N_c}-3 N_c & 0 & 6\\ \vspace*{0.1cm}
 \frac{3}{2}-\frac{3}{2 N_c^2} & 0 & -\frac{3}{N_c} & 0\\ \vspace*{0.1cm}
 0 & \frac{3}{2}-\frac{3}{2 N_c^2} & 0 & -\frac{3}{N_c}\\
\end{array}
\right),
\label{eq:lifetimes_ADM_QtilQtil}
\end{equation}
\begin{equation}
 \tilde{\gamma}_{\tilde{Q}\tilde{E}}^{(0)} = \left(
\begin{array}{ccccc}
 0 & 0 & -\frac{1}{2} & 0\\ \vspace*{0.1cm}
 0 & 0 & 0 & -\frac{1}{2}\\ \vspace*{0.1cm}
 \frac{1}{8 N_c^2}-\frac{1}{8} & 0 & \frac{1}{2 N_c}-\frac{N_c}{4} & 0\\ \vspace*{0.1cm}
 0 & \frac{1}{8 N_c^2}-\frac{1}{8} & 0 & \frac{1}{2 N_c}-\frac{N_c}{4}\\
\end{array}
\right).
\label{eq:lifetimes_ADM_QtilEtil}
\end{equation}
Our result~\eqref{eq:lifetimes_ADM_QQ} is in agreement with~\cite{Ciuchini:1997bw,Buras:2000if} 
and~\eqref{eq:lifetimes_ADM_QtilQtil} reproduces the result of \cite{Neubert:1996we}.\footnote{ 
Note that \cite{Neubert:1996we} contains a misprint that has been identified in~\cite{Ciuchini:2001vx}.} 
The results~\eqref{eq:lifetimes_ADM_QE} and~\eqref{eq:lifetimes_ADM_QtilEtil} are new. 
The matching coefficients read 
\begin{equation}
 C_{Q_i\tilde{Q}_j}^{(0)} = \delta_{ij}, 
\end{equation}
at LO and 
\begin{equation}
 C_{Q_i\tilde{Q}_j}^{(1)} = \left(
\begin{array}{cccc}
 -4 L_\mu-\frac{32}{3} & \frac{16}{3} & -\frac{a_1}{4}-3 L_\mu-13 & -2 \\
 0 & 4 L_\mu+\frac{16}{3} & -\frac{3}{2} & -\frac{a_2}{4}+3 L_\mu-1 \\
 -\frac{a_1}{18}-\frac{2 L_\mu}{3}-\frac{26}{9} & -\frac{4}{9} & -\frac{7 a_1}{24}+\frac{3 L_\mu}{2}+\frac{7}{6} & -3 \\
 -\frac{1}{3} & -\frac{a_2}{18}+\frac{2 L_\mu}{3}-\frac{2}{9} & -\frac{1}{4} & -\frac{7 a_2}{24}-\frac{3 L_\mu}{2}-\frac{29}{6} \\
\end{array}
\right), 
\end{equation}
at NLO where we have set $N_c=3$ for brevity.

\newpage
\section{Inputs and detailed overview of uncertainties\label{sec:Uncertainties}}

\begin{longtable}[h!]{|c|cc|}
\hline  
\rule{0pt}{3ex} Parameter                               & \hspace{2cm}Value\hspace{2cm}       & Source \\ \hline
\rule{0pt}{3ex} $\overline{m}_b(\overline{m}_b)$        & $(4.203_{-0.034}^{+0.016})$ GeV     & \cite{Beneke:2014pta,Beneke:2016oox} \\ 
\rule{0pt}{3ex} $m_b^\text{PS}(2\text{ GeV})$           & $(4.532_{-0.039}^{+0.013})$ GeV     & \cite{Beneke:2014pta,Beneke:2016oox} \\ 
\rule{0pt}{3ex} $m_b^\text{1S}$                         & $(4.66_{-0.03}^{+0.04})$ GeV        & \cite{Olive:2016xmw} \\ 
\rule{0pt}{3ex} $m_b^\text{kin}(1\text{ GeV})$          & $(4.553\pm0.020)$ GeV               & \cite{Alberti:2014yda} \\ 
\rule{0pt}{3ex} $\overline{m}_c(\overline{m}_c)$        & $(1.279\pm0.013)$ GeV               & \cite{Chetyrkin:2009fv} \\
\rule{0pt}{3ex} $\alpha_s(M_Z)$                         & $0.1181\pm0.0011$                   & \cite{Olive:2016xmw} \\ 
\rule{0pt}{3ex} $V_{us}$                                & $0.2248\pm0.0006$                   & \cite{Olive:2016xmw} \\ 
\rule{0pt}{3ex} $V_{ub}$                                & $0.00409\pm0.00039$                 & \cite{Olive:2016xmw} \\ 
\rule{0pt}{3ex} $V_{cb}$                                & $0.0405\pm0.0015$                   & \cite{Olive:2016xmw} \\ 
\rule{0pt}{3ex} $\gamma$                                & $(73.2_{-7.0}^{+6.3})^\circ$        & \cite{Olive:2016xmw} \\ 
\rule{0pt}{3ex} $f_{B}$                                 & $(189\pm4)$ MeV                     & \cite{Olive:2016xmw}\footnote{We take the mean of $f_{B^+}$ and $f_{B^0}$.} \\ 
\rule{0pt}{3ex} $f_{B_s}$                               & $(227.2\pm3.4)$ MeV                 & \cite{Olive:2016xmw} \\ 
\rule{0pt}{3ex} $f_{D}$                                 & $(203.7\pm4.8)$ MeV                 & \cite{Olive:2016xmw}\footnote{We use the 'experimental' value instead of the lattice average, since the former is in significantly better agreement with sum rule results~\cite{Wang:2015mxa,Gelhausen:2013wia,Narison:2012xy,Lucha:2011zp}.}\\
\hline
\caption{Input values for parameters.\label{tab:inputs}}
\end{longtable}

\begin{longtable}[h!]{|l|ccccccc|}
\hline \rule{0pt}{3ex} $\Delta B = 2$ & $\Lbar$              & intrinsic SR & condensates & $\mu_\rho$           & $1/m_b$    & $\mu_m$              & $a_i$\\ \hline
\rule{0pt}{3ex} $\overline{B}_{Q_1}$  & $_{-0.002}^{+0.001}$ & $\pm0.018$   & $\pm0.004$  & $_{-0.022}^{+0.011}$ & $\pm0.010$ & $_{-0.039}^{+0.045}$ & $_{-0.007}^{+0.007}$\\ 
\rule{0pt}{3ex} $\overline{B}_{Q_2}$  & $_{-0.017}^{+0.014}$ & $\mp0.020$   & $\pm0.004$  & $_{-0.019}^{+0.012}$ & $\pm0.010$ & $_{-0.062}^{+0.071}$ & $_{-0.015}^{+0.015}$\\ 
\rule{0pt}{3ex} $\overline{B}_{Q_3}$  & $_{-0.074}^{+0.060}$ & $\pm0.107$   & $\pm0.023$  & $_{-0.008}^{+0.016}$ & $\pm0.010$ & $_{-0.069}^{+0.086}$ & $_{-0.052}^{+0.053}$\\ 
\rule{0pt}{3ex} $\overline{B}_{Q_4}$  & $_{-0.006}^{+0.007}$ & $\pm0.021$   & $\pm0.011$  & $_{-0.003}^{+0.003}$ & $\pm0.010$ & $_{-0.079}^{+0.088}$ & $_{-0.006}^{+0.005}$\\ 
\rule{0pt}{3ex} $\overline{B}_{Q_5}$  & $_{-0.015}^{+0.019}$ & $\pm0.018$   & $\pm0.009$  & $_{-0.006}^{+0.004}$ & $\pm0.010$ & $_{-0.068}^{+0.077}$ & $_{-0.012}^{+0.012}$\\
\hline
\caption{Individual errors for the Bag parameters of the $\Delta B=2$ matrix elements.\label{tab:details_DelB2}}
\end{longtable}

\begin{longtable}[h!]{|l|ccccccc|}
\hline \rule{0pt}{3ex} $\Delta B = 0$     & $\Lbar$              & intrinsic SR & condensates & $\mu_\rho$           & $1/m_b$    & $\mu_m$              & $a_i$\\ \hline
\rule{0pt}{3ex} $\overline{B}_{1}$        & $_{-0.002}^{+0.003}$ & $\pm0.019$   & $\pm0.002$  & $_{-0.002}^{+0.002}$ & $\pm0.010$ & $_{-0.052}^{+0.060}$ & $_{-0.003}^{+0.002}$\\ 
\rule{0pt}{3ex} $\overline{B}_{2}$        & $_{-0.001}^{+0.001}$ & $\mp0.020$   & $\pm0.002$  & $_{-0.001}^{+0.000}$ & $\pm0.010$ & $_{-0.076}^{+0.084}$ & $_{-0.002}^{+0.001}$\\ 
\rule{0pt}{3ex} $\overline{\epsilon}_{1}$ & $_{-0.007}^{+0.006}$ & $\pm0.022$   & $\pm0.003$  & $_{-0.003}^{+0.003}$ & $\pm0.010$ & $_{-0.012}^{+0.010}$ & $_{-0.007}^{+0.006}$\\ 
\rule{0pt}{3ex} $\overline{\epsilon}_{2}$ & $_{-0.006}^{+0.005}$ & $\pm0.017$   & $\pm0.003$  & $_{-0.001}^{+0.002}$ & $\pm0.010$ & $_{-0.002}^{+0.001}$ & $_{-0.004}^{+0.003}$\\ 
\hline
\caption{Individual errors for the Bag parameters of the $\Delta B=0$ matrix elements.\label{tab:details_DelB0}}
\end{longtable}

\renewcommand{\footnoterule}{}

\begin{longtable}[h!]{|l|ccccccc|}
\hline \rule{0pt}{3ex} $\Delta C = 2$ & $\Lbar$              & intrinsic SR & condensates & $\mu_\rho$           & $1/m_c$    & $\mu_m$              & $a_i$\\ \hline
\rule{0pt}{3ex} $\overline{B}_{Q_1}$  & $_{-0.002}^{+0.001}$ & $\pm0.013$   & $\pm0.003$  & $_{-0.021}^{+0.009}$ & $\pm0.030$ & $_{-0.021}^{+0.039}$ & $\pm0.003$\\ 
\rule{0pt}{3ex} $\overline{B}_{Q_2}$  & $_{-0.014}^{+0.011}$ & $\mp0.015$   & $\pm0.003$  & $_{-0.016}^{+0.010}$ & $\pm0.030$ & $_{-0.050}^{+0.092}$ & $\pm0.012$\\ 
\rule{0pt}{3ex} $\overline{B}_{Q_3}$  & $_{-0.045}^{+0.037}$ & $\pm0.059$   & $\pm0.013$  & $_{-0.016}^{+0.016}$ & $\pm0.030$ & $_{-0.059}^{+0.116}$ & $\pm0.016$\\ 
\rule{0pt}{3ex} $\overline{B}_{Q_4}$  & $_{-0.005}^{+0.006}$ & $\pm0.017$   & $\pm0.009$  & $_{-0.003}^{+0.003}$ & $\pm0.030$ & $_{-0.071}^{+0.131}$ & $\pm0.004$\\ 
\rule{0pt}{3ex} $\overline{B}_{Q_5}$  & $_{-0.012}^{+0.014}$ & $\pm0.014$   & $\pm0.007$  & $_{-0.005}^{+0.004}$ & $\pm0.030$ & $_{-0.069}^{+0.127}$ & $\pm0.004$\\
\hline
\caption{Individual errors for the Bag parameters of the $\Delta C=2$ matrix elements.\label{tab:details_DelC2}}
\end{longtable}

\newpage

\begin{longtable}[h!]{|l|ccccccc|}
\hline \rule{0pt}{3ex} $\Delta C = 0$     & $\Lbar$              & intrinsic SR & condensates & $\mu_\rho$           & $1/m_c$    & $\mu_m$              & $a_i$\\ \hline
\rule{0pt}{3ex} $\overline{B}_{1}$        & $_{-0.003}^{+0.004}$ & $\pm0.017$   & $\pm0.002$  & $_{-0.002}^{+0.002}$ & $\pm0.030$ & $_{-0.037}^{+0.068}$ & $_{-0.005}^{+0.003}$\\ 
\rule{0pt}{3ex} $\overline{B}_{2}$        & $_{-0.000}^{+0.001}$ & $\mp0.015$   & $\pm0.001$  & $_{-0.000}^{+0.000}$ & $\pm0.030$ & $_{-0.065}^{+0.120}$ & $_{-0.001}^{+0.000}$\\ 
\rule{0pt}{3ex} $\overline{\epsilon}_{1}$ & $_{-0.008}^{+0.007}$ & $\pm0.024$   & $\pm0.004$  & $_{-0.004}^{+0.003}$ & $\pm0.030$ & $_{-0.022}^{+0.012}$ & $_{-0.008}^{+0.006}$\\ 
\rule{0pt}{3ex} $\overline{\epsilon}_{2}$ & $_{-0.004}^{+0.003}$ & $\pm0.011$   & $\pm0.002$  & $_{-0.001}^{+0.001}$ & $\pm0.030$ & $_{-0.000}^{+0.000}$ & $_{-0.002}^{+0.001}$\\ 
\hline
\caption{Individual errors for the Bag parameters of the $\Delta C=0$ matrix elements.\label{tab:details_DelC0}}
\end{longtable}

\begin{longtable}[h!]{|l|ccc|}
\hline \rule{0pt}{3ex}                    & $\Delta M_s^\text{SM}$ [ps$^{-1}$] & $\Delta \Gamma_s^\text{PS}$ [ps$^{-1}$] & $a_\text{sl}^{s,\,\text{PS}}$ [$10^{-5}$] \\ \hline
\rule{0pt}{3ex} $\overline{B}_{Q_ 1}$      & $\pm1.0$                           & $\pm0.005$                              & $\pm0.01$ \\ 
\rule{0pt}{3ex} $\overline{B}_{Q_ 3}$      & $\pm0.0$                           & $\pm0.005$                              & $\pm0.02$ \\ 
\rule{0pt}{3ex} $\overline{B}_{R_ 0}$      & $\pm0.0$                           & $\pm0.003$                              & $\pm0.00$ \\ 
\rule{0pt}{3ex} $\overline{B}_{R_ 1}$      & $\pm0.0$                           & $\pm0.000$                              & $\pm0.00$ \\
\rule{0pt}{3ex} $\overline{B}_{R_ 1'}$     & $\pm0.0$                           & $\pm0.000$                              & $\pm0.00$ \\
\rule{0pt}{3ex} $\overline{B}_{R_ 2}$      & $\pm0.0$                           & $\pm0.016$                              & $\pm0.00$ \\
\rule{0pt}{3ex} $\overline{B}_{R_ 3}$      & $\pm0.0$                           & $\pm0.000$                              & $\pm0.02$ \\
\rule{0pt}{3ex} $\overline{B}_{R_ 3'}$     & $\pm0.0$                           & $\pm0.000$                              & $\pm0.05$ \\
\rule{0pt}{3ex} $f_{B_s}$                 & $\pm0.5$                           & $\pm0.002$                              & $\pm0.00$ \\ 
\rule{0pt}{3ex} $\mu_1$                   & $\pm0.0$                           & $_ {-0.018}^{+0.007}$                    & $_ {-0.09}^{+0.04}$ \\ 
\rule{0pt}{3ex} $\mu_2$                   & $\pm0.1$                           & $_ {-0.002}^{+0.000}$                    & $_ {-0.01}^{+0.02}$ \\ 
\rule{0pt}{3ex} $m_b$                     & $\pm0.0$                           & $_ {-0.001}^{+0.000}$                    & $_ {-0.01}^{+0.02}$ \\ 
\rule{0pt}{3ex} $m_c$                     & $\pm0.0$                           & $_ {-0.001}^{+0.000}$                    & $\pm0.07$ \\ 
\rule{0pt}{3ex} $\alpha_s$                & $\pm0.0$                           & $\pm0.000$                              & $\pm0.04$ \\ 
\rule{0pt}{3ex} CKM                       & $_ {-1.2}^{+1.3}$                   & $\pm0.006$                              & $_ {-0.24}^{+0.23}$ \\ 
\hline
\caption{Individual errors for the $B_s$ mixing observables.\label{tab:details_Bs_mixing}}
\end{longtable}

\newpage

\begin{table}[h!]
 \centering
 \begin{tabular}
{|l|ccc|}
\hline \rule{0pt}{3ex}                    & $\Delta M_d^\text{SM}$ [ps$^{-1}$] & $\Delta \Gamma_d^\text{PS}$ [10$^{-3}$\,ps$^{-1}$] & $a_\text{sl}^{d,\,\text{PS}}$ [$10^{-4}$] \\ \hline
\rule{0pt}{3ex} $\overline{B}_{Q_1}$      & $\pm0.03$                          & $\pm0.16$                                          & $\pm0.02$ \\ 
\rule{0pt}{3ex} $\overline{B}_{Q_3}$      & $\pm0.00$                          & $_{-0.16}^{+0.17}$                                 & $\pm0.03$ \\ 
\rule{0pt}{3ex} $\overline{B}_{R_0}$      & $\pm0.00$                          & $\pm0.11$                                          & $\pm0.01$ \\ 
\rule{0pt}{3ex} $\overline{B}_{R_1}$      & $\pm0.00$                          & $\pm0.01$                                          & $\pm0.00$ \\
\rule{0pt}{3ex} $\overline{B}_{R_1'}$     & $\pm0.00$                          & $\pm0.01$                                          & $\pm0.00$ \\
\rule{0pt}{3ex} $\overline{B}_{R_2}$      & $\pm0.00$                          & $\pm0.54$                                          & $\pm0.00$ \\
\rule{0pt}{3ex} $\overline{B}_{R_3}$      & $\pm0.00$                          & $\pm0.00$                                          & $\pm0.04$ \\
\rule{0pt}{3ex} $\overline{B}_{R_3'}$     & $\pm0.00$                          & $\pm0.01$                                          & $\pm0.10$ \\
\rule{0pt}{3ex} $f_{B}$                   & $\pm0.02$                          & $\pm0.11$                                          & $\pm0.00$ \\ 
\rule{0pt}{3ex} $\mu_1$                   & $\pm0.00$                          & $_{-0.62}^{+0.24}$                                 & $_{-0.08}^{+0.19}$ \\ 
\rule{0pt}{3ex} $\mu_2$                   & $\pm0.00$                          & $_{-0.08}^{+0.00}$                                 & $_{-0.03}^{+0.01}$ \\ 
\rule{0pt}{3ex} $m_b$                     & $\pm0.00$                          & $_{-0.03}^{+0.01}$                                 & $_{-0.03}^{+0.01}$ \\ 
\rule{0pt}{3ex} $m_c$                     & $\pm0.00$                          & $\pm0.02$                                          & $\pm0.14$ \\ 
\rule{0pt}{3ex} $\alpha_s$                & $\pm0.00$                          & $\pm0.01$                                          & $_{-0.08}^{+0.09}$ \\ 
\rule{0pt}{3ex} CKM                       & $\pm0.07$                          & $_{-0.37}^{+0.36}$                                 & $\pm0.49$\\ 
\hline
\end{tabular}
\caption{Individual errors for the $B_d$ mixing observables.\label{tab:details_Bd_mixing}}
\end{table}

\begin{table}[h!]
 \centering
\begin{tabular}{|cccccccc|}
\hline \rule{0pt}{3ex} $\overline{B}_1$ & $\overline{B}_2$ & $\overline{\epsilon}_1$ & $\overline{\epsilon}_2$ & $\rho_3$   & $\rho_4$   & $\sigma_3$ & $\sigma_4$\\ \hline
\rule{0pt}{3ex}        $\pm0.002$       & $\pm0.000$       & $_{-0.015}^{+0.016}$    & $\pm0.004$              & $\pm0.001$ & $\pm0.000$ & $\pm0.013$ & $\pm0.000$ \\ \hline
\hline \rule{0pt}{3ex} $f_B$                & $\mu_1$              & $\mu_0$              & $m_b$                & $m_c$      & $\alpha_s$ & CKM        & \\ \hline
\rule{0pt}{3ex}        $_{-0.003}^{+0.004}$ & $_{-0.013}^{+0.000}$ & $_{-0.006}^{+0.000}$ & $_{-0.001}^{+0.000}$ & $\pm0.000$ & $\pm0.002$ & $\pm0.006$ &  \\
\hline
\end{tabular}
\caption{Individual errors for the ratio $\tau(B^+)/\tau(B^0)$ in the PS mass scheme.\label{tab:details_BpB0}}
\end{table}

\begin{table}[h!]
 \centering
\begin{tabular}{|cccccccc|}
\hline \rule{0pt}{3ex} $\overline{B}_1$   & $\overline{B}_2$ & $\overline{\epsilon}_1$ & $\overline{\epsilon}_2$ & $\rho_3$  & $\rho_4$  & $\sigma_3$ & $\sigma_4$\\ \hline
\rule{0pt}{3ex}        $_{-0.05}^{+0.07}$ & $\pm0.00$        & $_{-0.47}^{+0.52}$      & $\pm0.017$              & $\pm0.05$ & $\pm0.00$ & $\pm0.46$  & $\pm0.00$ \\ \hline
\hline \rule{0pt}{3ex} $f_B$     & $\mu_1$            & $\mu_0$            & $m_c$     & $m_s$     & $\alpha_s$        & CKM       & \\ \hline
\rule{0pt}{3ex}        $\pm0.08$ & $_{-0.40}^{+0.07}$ & $_{-0.21}^{+0.08}$ & $\pm0.08$ & $\pm0.00$ & $_{0.06}^{+0.07}$ & $\pm0.00$ &  \\
\hline
\end{tabular}
\caption{Individual errors for the ratio $\tau(D^+)/\tau(D^0)$ in the PS mass scheme.\label{tab:details_DpD0}}
\end{table}

\end{document}